\documentclass[ALICE,manyauthors]{cernphprep}

\usepackage{rotating}
\usepackage[normalem]{ulem}
\usepackage{amssymb,amsmath,amsfonts}
\usepackage{graphicx}
\usepackage{subfigure}
\usepackage{amssymb}
\usepackage{epstopdf}
\usepackage{cite}
\usepackage{multirow}
\usepackage{color}
\usepackage{amsmath,bm}
\usepackage{amsbsy}
\usepackage{ulem}
\usepackage{siunitx}

\usepackage{hyperref}
\usepackage{datetime}
\usdate
\usepackage{lineno}
\usepackage[toc,page]{appendix}
\usepackage[section] {placeins}

\newcommand{\CommentBlock}[1]{}

\newcommand{\pp}{pp}

\newcommand{\sqrts}{\ensuremath{\sqrt{s}}}

\newcommand{\gev}{\ensuremath{\mathrm{GeV/}c}}

\newcommand{\kT}{\ensuremath{k_\mathrm{T}}}
\newcommand{\antikT}{anti-\ensuremath{k_\mathrm{T}}}

\begin{document}

\begin{titlepage}
\PHnumber{201}                 
\PHdate{16 July}              
\PHyear{2018}              

\title{Medium modification of the shape of small-radius jets in central Pb--Pb collisions at $\mathbf{\sqrt{s_{\mathrm {NN}}}} = 2.76\,\rm{TeV}$}
\ShortTitle{Small-radius jet shapes in pp and Pb--Pb collisions at ALICE } 

\Collaboration{ALICE Collaboration%
         \thanks{See Appendix~\ref*{app:collab} for the list of collaboration
                      members}}
\ShortAuthor{ALICE Collaboration}      

\begin{abstract}
 We present the measurement of a new set of jet shape observables
 for track-based jets in central Pb--Pb collisions at $\ensuremath{\sqrt{s_{\mathrm{NN}}}} = 2.76$\,TeV. The set of jet shapes includes the first radial moment or angularity, $g$; the momentum dispersion,
 $p_{\rm T}D$; and the difference between the leading and sub-leading
 constituent track transverse momentum, $LeSub$. These observables provide complementary information on the
 jet fragmentation and can constrain different aspects of the theoretical description of jet-medium interactions.
 \noindent The jet shapes were measured for a small resolution parameter $R = 0.2$ and were fully corrected to particle level.
 The observed jet shape modifications indicate that in-medium fragmentation is
 harder and more collimated than vacuum fragmentation as obtained by
 PYTHIA calculations, which were validated with the
 measurements of the jet shapes in proton-proton collisions
 at $\sqrts = 7$\,TeV.
 The comparison of the measured distributions to
 templates for quark and gluon-initiated jets indicates that in-medium fragmentation resembles that of quark jets in vacuum.
 We further argue that the observed modifications are not consistent with a totally coherent energy loss picture where the jet loses energy as a single colour charge, suggesting that the medium resolves the jet structure at the angular scales probed by our measurements ($R=0.2$).
 Furthermore, we observe that small-$R$ jets can help to isolate purely energy loss effects from other effects that contribute to the modifications of the jet shower in medium such as the correlated background or medium response. 
\end{abstract}

\maketitle


\end{titlepage}
 
\setcounter{page}{2}
\section{Introduction}
\label{sect:intro}

The objective of the heavy-ion jet physics program at RHIC and LHC is to understand the
behaviour of QCD matter at the limit of high energy density and temperature by studying
the dynamics of jet-medium interactions. Jet physics in heavy-ion
collisions is a multiscale problem. Hard scales govern the elementary
scattering and the subsequent branching process down to
non-perturbative scales, in the vacuum as well as in the medium. Soft scales, of the order of the temperature of the medium, characterise the interactions of soft partons produced in the shower with the strongly coupled medium. Soft scales also rule hadronisation, which is expected to take place in vacuum for sufficiently energetic probes. A detailed discussion of the different processes contributing to the jet shower evolution in medium and their onset scales can be found in Ref.~\cite{Kurkela:2014tla}.
The interplay between these processes can lead to modifications of the
longitudinal and transverse distributions of the constituents of the
jet with respect to jet fragmentation in vacuum. 
These jet structure modifications can be investigated with jet shape observables and have the potential to constrain the dynamics of jet energy loss in medium, the role of colour coherence~\cite{CasalderreySolana:2012ef}, and fundamental medium properties like temperature, density or the evolution of the medium degrees of freedom with the resolution scale \cite{D'Eramo:2012jh}.

The jet shape observables measured so far in heavy-ion collisions at the LHC can be classified into three groups: inclusive, jet-by-jet shapes using constituents information, and jet shapes using the clustering history. 
The first group consists on inclusive observables that measure intra or inter-jet distributions. The ratios of jet yields with
different resolution parameter $R$ are an example. Such ratios are infrared and collinear (IRC) safe \cite{Salam:2009jx} and are sensitive to the transverse energy profile of the jets \cite{Abelev:2013fn,Chatrchyan:2014gia,Dasgupta:2016bnd}. ATLAS measured central to peripheral inclusive jet yield ratios for different jet radii up to $R=0.5$ showing differences of the order of $30\%$ at $p_{\rm{T,jet}}< 100$\,GeV/$c$ \cite{Aad:2012vca}, which indicate energy redistribution within the jet in medium relative to vacuum. 
In ALICE, such ratios were measured for inclusive and semi-inclusive
samples of jets recoiling from high-$p_{\rm T}$
hadrons~\cite{Abelev:2013kqa, Adam:2015doa}. In the case of recoil
jets, larger $R$ were accessible and the results showed no indication
of medium modifications when changing the jet resolution up to $R
=0.5$. ALICE and ATLAS measurements are characterised by different jet selections and different minimum constituent cutoffs.
 Another example of shapes belonging to this category are the fragmentation fuctions \cite{Aaboud:2017bzv, Chatrchyan:2014ava}. The fragmentation functions give information on the longitudinal
 share of energy within the jet. The experimental results show an
 enhancement of the low and high-$z$ component and a depletion at
 intermediate $z$ in Pb--Pb relative to pp collisions, where $z$ is
 the fraction of the jet momentum carried by the particles in the jet
 \cite{Aaboud:2017bzv, Chatrchyan:2014ava}. The modifications are
 small and they were quantified as an excess of approximately $0.9$ particles
 at low momentum, in the difference between the integrals of the
 fragmentation functions in Pb--Pb and pp collisions. 

In order to probe the jet shape at large angles relative to the jet
 axis, two observables were designed. The CMS missing $p_{\rm T}$
 method~\cite{Khachatryan:2015lha} considers the projection of all
 particle  momentum vectors in the event onto the axis of a selected 
 dijet pair. This method is insensitive to the uncorrelated background, and
 particles correlated with the dijet reveal that momentum
 balance of the system is totally recovered only by very soft particles ($p_{\mathrm{T}} \le
 1$\,GeV/$c$) at large angles ($\Delta R_{\rm jet} > 0.8$). Jet-track
 angular correlations~\cite{Khachatryan:2016erx} explore the large-angle component differentially with similar conclusions. Similarly, the jet profile~\cite{Chatrchyan:2013kwa} measures the radial distribution of energy relative to the jet axis. The results indicate an enhancement of momentum relative to pp collisions at distances to the jet axis $\Delta R_{\rm jet} \gtrsim 0.3$. This enhancement is accompanied by a reduction of momentum at short distances to the jet axis $0.1 < \Delta R_{\rm jet} < 0.2$.

 The second group of shape observables are the jet shapes built as a jet-by-jet function of the jet constituent 4-momenta.
 The jet mass \cite{Acharya:2017goa} is an example. The jet mass is related to the
 virtuality of the parton that originated the jet. It increases
 with large-angle soft particle emission. The ALICE measurement of the
 jet mass in heavy-ion collisions~\cite{Acharya:2017goa} for jets of $R = 0.4$ showed a hint of reduction relative to the vacuum reference. Theoretical models show that energy loss effects reduce the jet mass while the medium response increases it, resulting in a mass that is shifted to higher values than what was found by ALICE results \cite{KunnawalkamElayavalli:2017hxo}.

 
The third category of jet shape observables uses the clustering history to select certain parts of the particle shower using
well-defined jet clustering techniques, for instance grooming~\cite{Butterworth:2008iy,Larkoski:2014wba}, to amplify or
suppress a region of the splitting phase space where medium-induced
effects are expected. Examples are the 2-subjetiness~\cite{Zardoshti:2017yiy} or the soft drop subjet momentum balance, $z_{\rm g}$ \cite{Sirunyan:2017bsd, Kauder:2017mhg}, designed to explore changes in the rate of 2-prong jets and the momentum balance of semi-hard subjets in heavy-ion collisions relative to pp collisions. New ideas and applications for this third category of jet shapes are being discussed in the literature for beyond Standard Model searches and QCD studies in pp as well as heavy-ion collisions.

The shapes analysed in this paper belong to the second category and
are described in detail in Section~\ref{sect:definition}. They probe
complementary aspects of the jet fragmentation such as the transverse
energy profile or the dispersion of the jet constituents transverse
momentum distribution. Our aim was to perform a systematic exploration
of the intrajet distributions to pose constraints on key aspects of the
theory of jet quenching. A clean connection to the theory was pursued
via the selection of observables that are well defined and calculable
from first principles in pQCD and via the full correction of the
observables to particle level. The considered small resolution $R=0.2$
and ALICE instrumental capabilities allowed us to obtain fully
corrected particle-level jet measurements, in a unique range at the LHC of low jet momentum and low constituent momentum cutoff of $0.15$ GeV/$c$. Our measurements give insight on whether the jet substructure is resolved by the medium at small angular scales and on the role of the medium response.

The rest of the paper is organized as follows: Section~\ref{sect:Dataset} presents the data sets and event selection used for the
analysis, Sections~\ref{sect:JetReco} and~\ref{sect:subtraction} describe the jet finding procedure and the underlying event
subtraction, while Sections~\ref{sect:response} and~\ref{sect:unfolding} 
present the response of the shapes to detector effects and background
fluctuations and the 2-dimensional unfolding procedure that
simultaneously corrects the shape and jet $p_{\rm T}$ distributions. Section~\ref{sect:sysuncert} describes the different contributions to the
systematic uncertainty and finally, Section~\ref{sect:results} presents the fully corrected results and their interpretation with comparisons to theoretical models. 

 \section{The set of jet shape observables}
 \label{sect:definition}
 \noindent In this analysis, we focus on three jet shape observables that probe complementary
aspects of the jet fragmentation, namely the first radial moment or angularity (or girth), $g$, the momentum dispersion, $p_{\mathrm{T}}D$, and the difference between the leading and sub-leading track transverse momentum, $LeSub$. 

\noindent The angularity is defined as
\begin{equation}
\label{eq:angu}
g = \sum_{i \in \rm jet} \frac{p_{\mathrm{T,}i}}{p_{\mathrm{T,jet}}} \Delta R_{\mathrm{jet},i},
\end{equation} 
where $p_{\mathrm{T},i}$ is the transverse momentum of the $i$-th constituent and
$\Delta R_{\mathrm{jet},i}$ is the distance in ($\eta$, $\varphi$) space between
constituent $i$ and the jet axis. This shape is sensitive to the radial
energy profile of the jet.

\noindent The momentum dispersion $p_{\mathrm{T}}D$ is defined as

\begin{equation}
\label{eq:ptd}
 p_{\rm T}D = \frac{\sqrt{\sum_{i \in \rm jet}  p_{\mathrm{T,}i}^{2}}} {\sum_{i \in \rm jet} {p_{\mathrm{T,}i}}}.
\end{equation} 
This shape measures the second moment of the constituent $p_{\rm T}$
distribution in the jet and is connected to how hard or soft the jet fragmentation is. For
example, in the extreme case of few constituents carrying a
large fraction of the jet momentum, $p_{\mathrm{T}}D$ will be close to 1, while in
the case of jets with a large number of constituents and softer
momentum, $p_{\mathrm{T}}D$ would end up closer to 0. 


The two previous shapes are related to the moments of the so-called
generalized angularities defined as:
$\lambda_{\beta}^{\kappa} = \sum_{i} (\frac{p_{\rm{T},\it{i}}}{p_{\rm{T,jet}}})^{\kappa} (\frac{\Delta R_{\rm{jet},\it{i}}}{R})^{\beta}$ ~\cite{Larkoski:2014pca}. The
number of jet constituents corresponds to ($\kappa$,$\beta$) $=$ (0,0), the square of $p_{\rm T} D$ corresponds to (2,0), the angularity $g$ corresponds to (1,1),
and the square of the mass scaled by the jet $p_{\rm T}$ is related to (1,2).

\noindent $LeSub$ is defined as the difference of the leading  track $p_{\rm T}$ ( $p_{\mathrm{T,track}}^{\rm{lead}}$) and sub-leading track $p_{\rm T}$ ($p_{\mathrm{T,track}}^{\rm{sublead}}$):

\begin{equation}
\label{eq:angu}
LeSub = p_{\mathrm{T,track}}^{\rm {lead}}-p_{\mathrm{T,track}}^{\rm {sublead}}.
\end{equation} 

\begin{figure}[h]
  \includegraphics[width=0.33\textwidth]{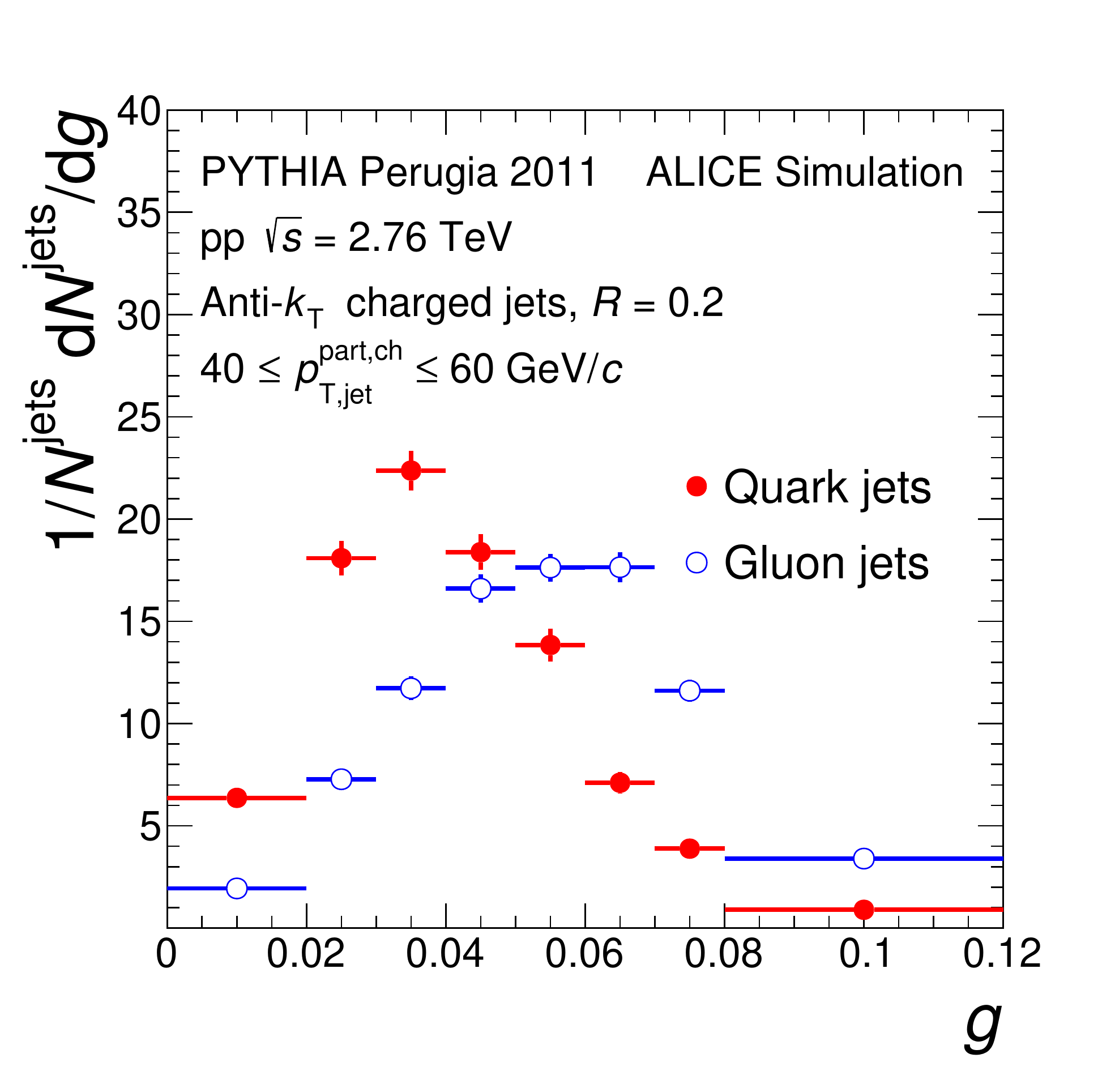}
  \includegraphics[width=0.33\textwidth]{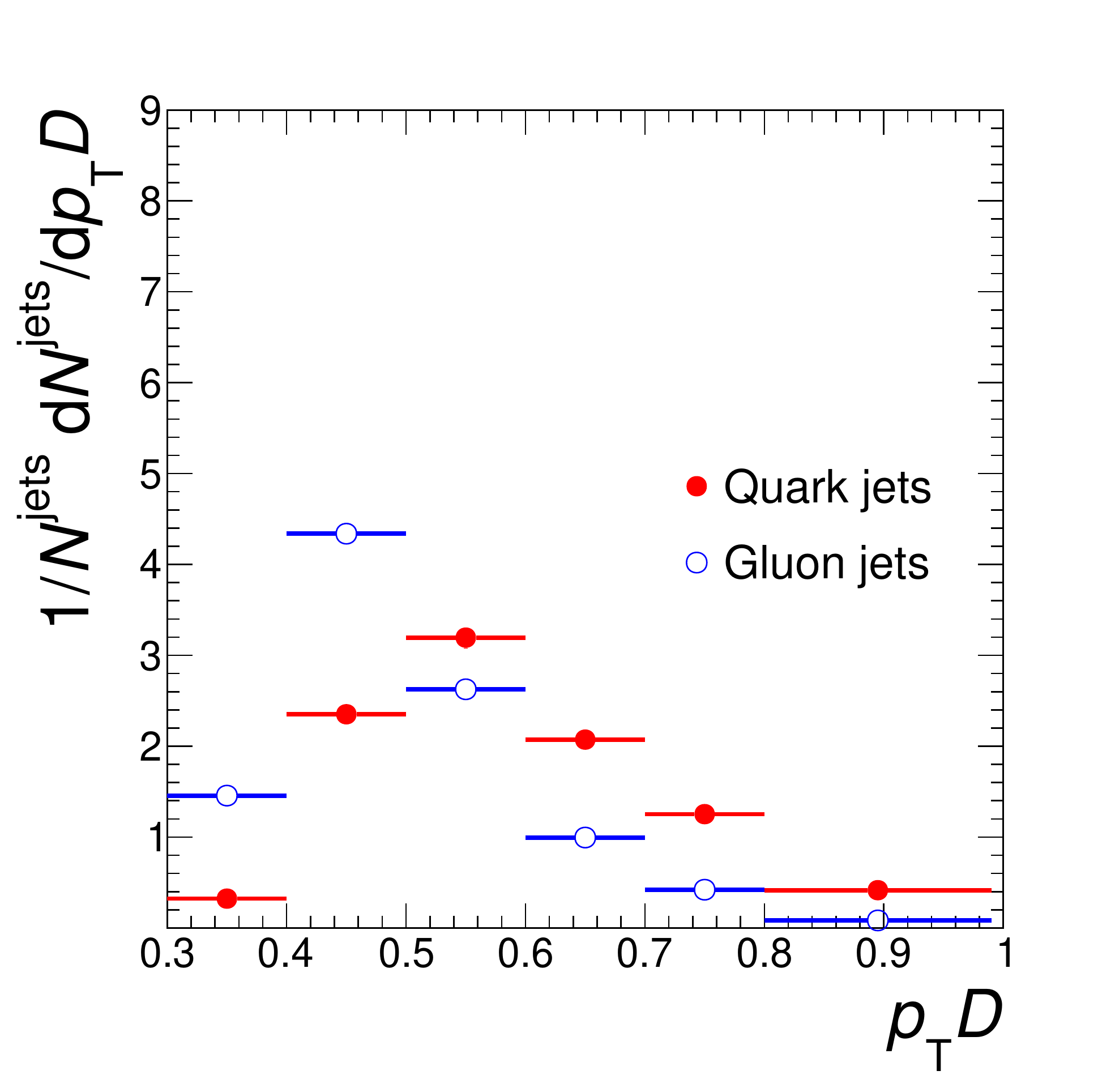}
\includegraphics[width=0.33\textwidth]{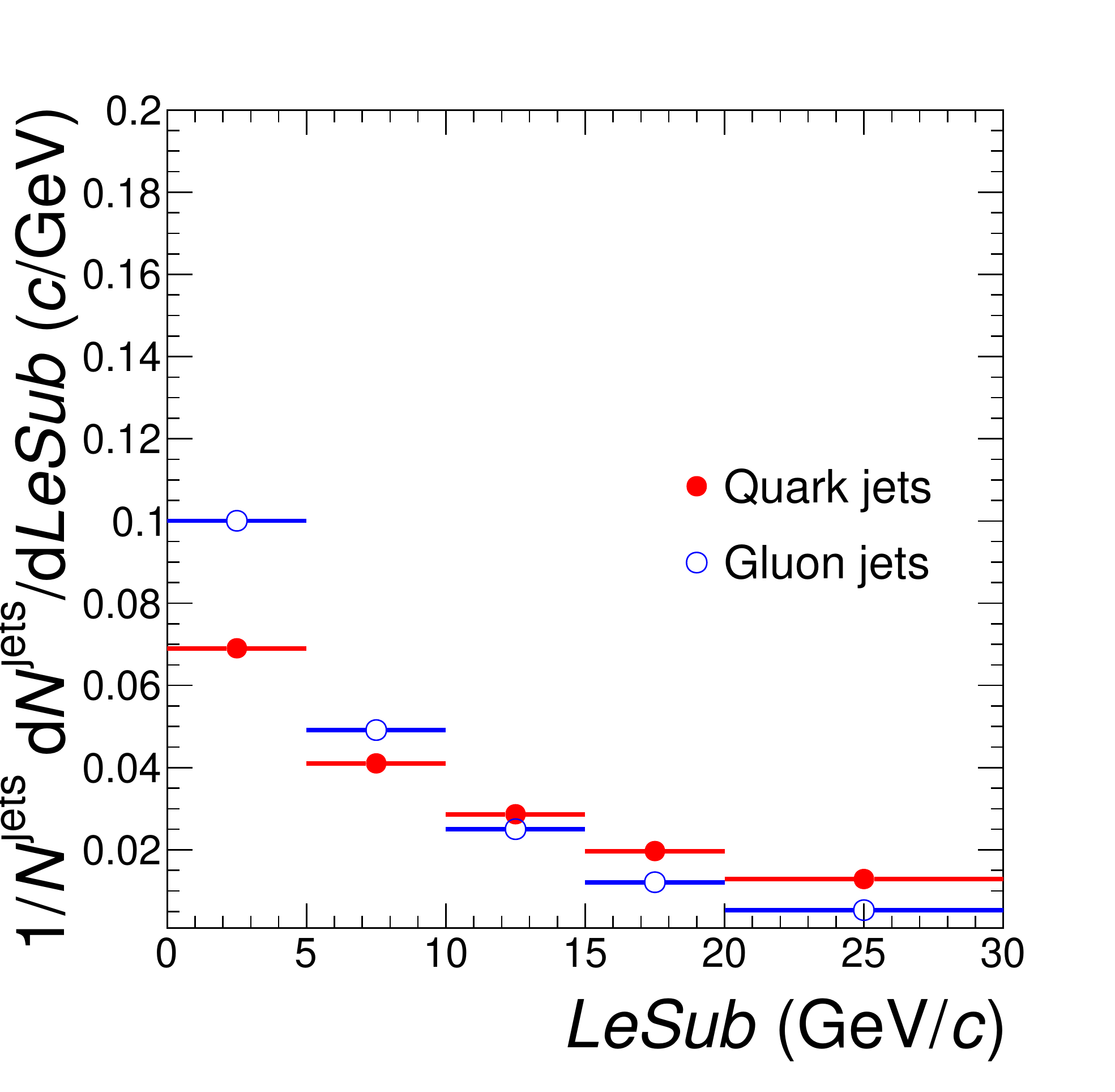}
\caption{$g$, $p_{\rm T}D$, and $LeSub$ for quark and gluon jets as
  obtained from PYTHIA Perugia 2011 simulations of pp collisions at $\sqrt{s}=2.76$\,TeV
  in the transverse momentum interval $40 \leq  p_{\mathrm{T,jet}}^{\rm{part,ch}} \leq 60$\,GeV/$c$.}
\label{fig:qgpythia}
\end{figure}

\begin{figure}[h]
  \includegraphics[width=0.33\textwidth]{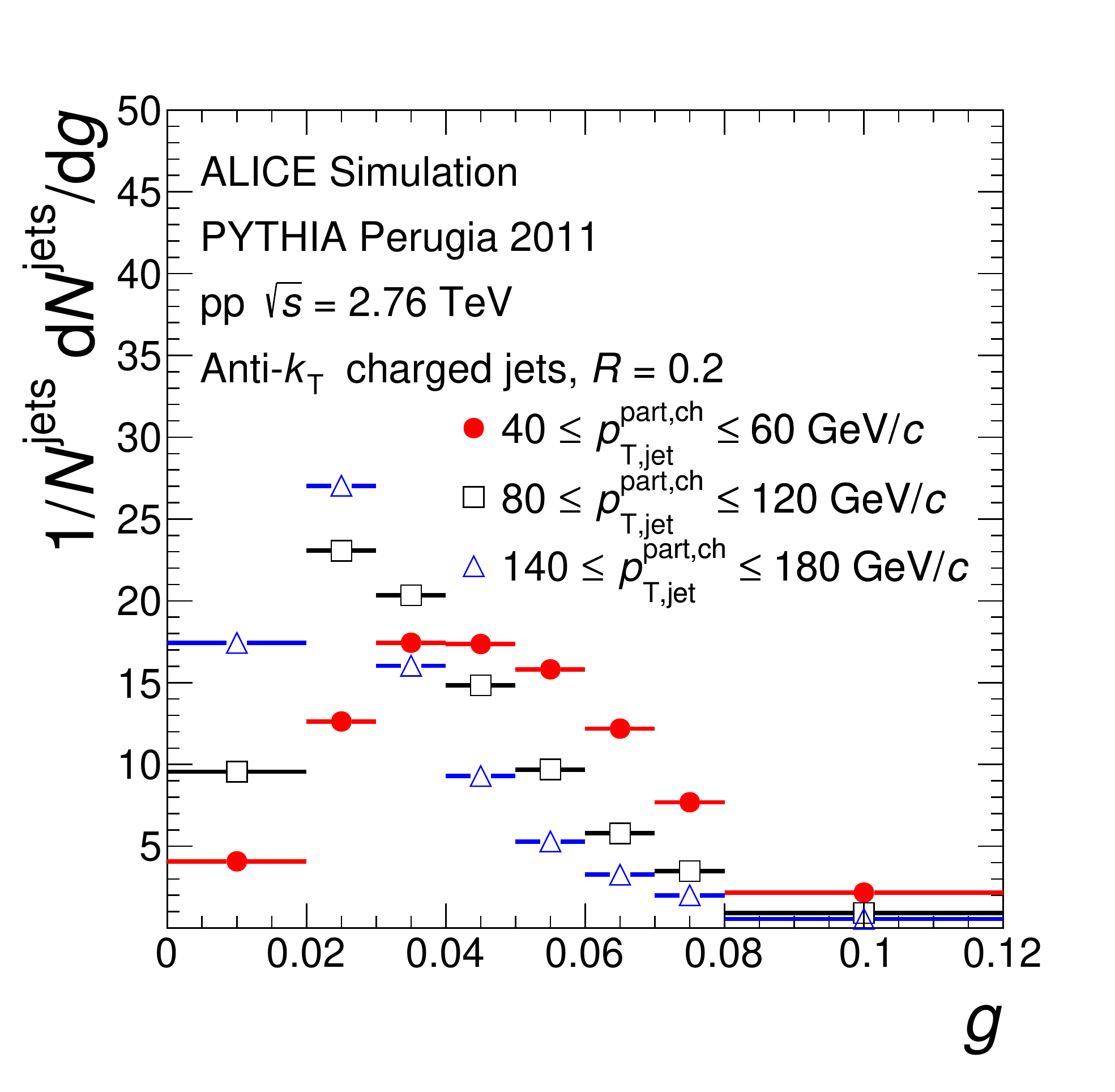}
  \includegraphics[width=0.33\textwidth]{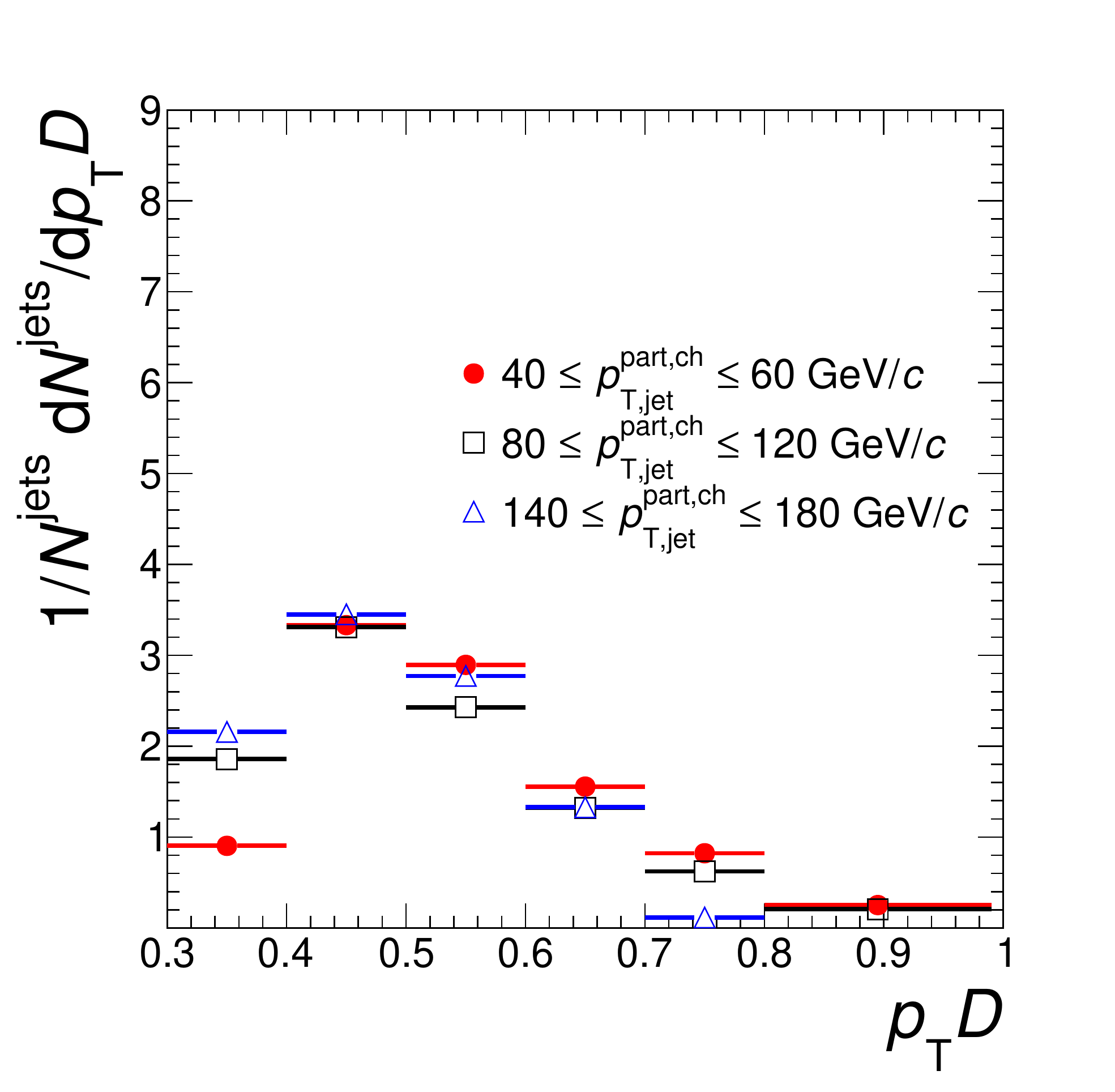}
\includegraphics[width=0.33\textwidth]{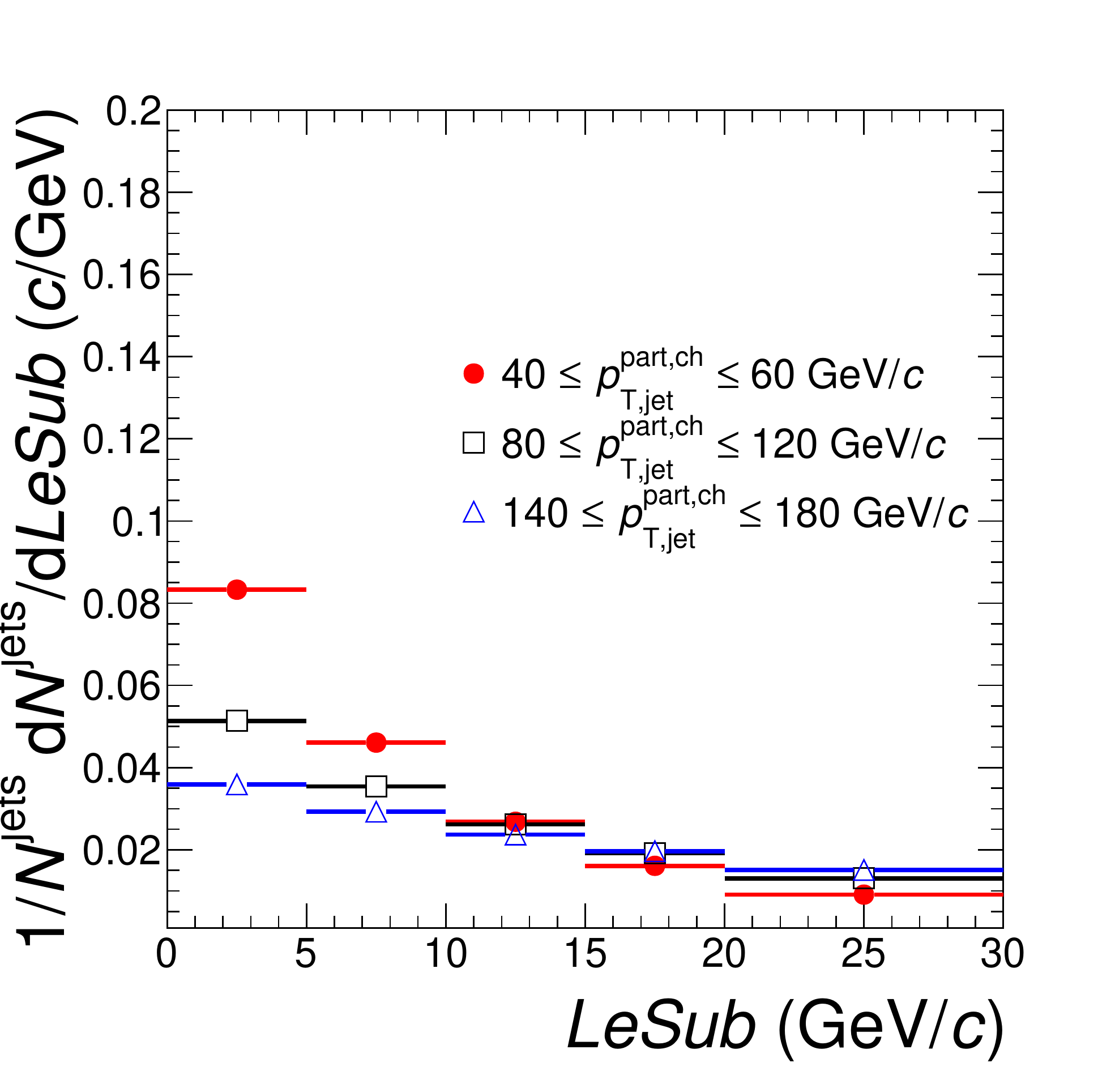}
\caption{$g$, $p_{\rm T}D$, and $LeSub$  as
  obtained from PYTHIA Perugia 2011 simulations of pp collisions at $\sqrt{s}=2.76$\,TeV
  for three different transverse momentum intervals.}
\label{fig:ptdeppythia}
\end{figure}

$LeSub$ is not an IRC-safe observable but shows robustness against contributions of soft background particles as we will discuss in
Section~\ref{sect:subtraction}.
In order to give an illustrative example for the sensitivity of these
observables to different types of jet fragmentation,
Fig.~\ref{fig:qgpythia} compares the behaviour of the shapes distributions for quark
and gluon initiated jets as obtained by
PYTHIA~\cite{Sjostrand:2006za} in pp collisions. At the same transverse momentum, gluon jets are broader and
produce more fragments with a softer momentum spectrum than quark jets. Consequently, their first radial moment ($g$) is on average higher, whereas the momentum dispersion ($p_{\rm T}D$) and $LeSub$ are lower. 
The momentum dependence of the three shapes in vacuum is illustrated in Fig.~\ref{fig:ptdeppythia}. As the jet momentum increases, the angularity and the $p_{\rm{T}}D$ decrease while $LeSub$ shifts to higher values. These changes are consistent with jets becoming narrower and with larger differences among constituents' transverse momentum at higher jet $p_{\rm{T}}$. 

\section{Data sets, event selection, and simulations}
\label{sect:Dataset}

The ALICE detector and its performance are described in
Refs.~\cite{Aamodt:2008zz,Abelev:2014ffa}. 

The data were taken during the 2011 LHC Pb--Pb run
at $\sqrt{s_{\rm NN}} = 2.76$\,TeV. 
This analysis uses events recorded with minimum-bias
(MB) triggers, based on the signal measured in the V0 scintillators
detectors that cover the full azimuth in the pseudo-rapidity intervals
$-3.7 < \eta < -1.7$ and $2.8 < \eta < 5.1$. The online information of the V0 detector was also used to
enhance the fraction of the 10\% most-central Pb--Pb
collisions. The online centrality selection has an efficiency of 100\% for the $0$--$7\%$ interval in
centrality percentile and drops to about 80\% efficiency for the $7$--$10\%$
interval.

Events are reconstructed offline as described in
Ref.~\cite{Abelev:2012hxa}. Charged tracks are measured in the ALICE
central barrel via the Inner Tracking
System (ITS), which consists of six cylindrical layers of silicon
detectors, and the Time Projection Chamber (TPC)  with acceptance
$|\eta|<0.9$ over the full azimuth. Accepted tracks were required to have $0.15<p_{\rm T}< 100$~\gev,
with at least 70 space-points and at
least 80\% of the geometrically findable space-points in the TPC.  To
account for the azimuthally non-uniform response of the ITS, in this dataset, two exclusive classes of tracks were used
\cite{Abelev:2014ffa}: tracks with Silicon Pixel Detector (SPD) hits
(70\% of all tracks in central Pb--Pb collisions and 95\% in \pp\
collisions) and, when the SPD information is not present, TPC tracks
with at least one hit in the ITS, with their trajectory refitted to the primary
vertex to improve the momentum resolution. The primary vertex was required to lie within 10\,cm of the
nominal center of the detector along the beam axis and within 1\,cm of
it in the transverse plane.  After offline event selection, the Pb--Pb
dataset consisted of 17M events in the $0$--$10\%$ centrality
percentile interval ($L_{\rm int} \approx \SI{21.3}{\micro b}^{-1}$).

The \pp\ collision data used to validate PYTHIA~\cite{Sjostrand:2006za} 
were recorded during the 2010 low-luminosity \pp\ run at $\sqrt{s} = 7$\,TeV with a MB trigger selection. The trigger configuration, offline event
selection and tracking are described in
Ref. \cite{ALICE:2014dla}. After the event selection, the \pp\ dataset
consisted of 168M events ($L_{\rm int} \approx \SI{2.5}{\nano b}^{-1}$). 

For central Pb--Pb collisions, the tracking efficiency is about $80\%$ for $p_{\rm T}>1$\,\gev, decreasing to $\approx 56\%$ at $p_{\rm T} = 0.15$\,\gev. The track momentum resolution is
around $1\%$ at $p_{\rm T}\, = 1$\,\gev\, and $\approx 2.5\%$ at $p_{\rm T}\, = 40$\,\gev. For \pp\ collisions,
the tracking efficiency is about $2$--$3\%$ higher than in central Pb--Pb
collisions. The track momentum resolution is about $1\%$ for reconstructed
tracks with $p_{\rm T}\, = 1$\,\gev\, and of the order of $4.1\%$ for $p_{\rm T}\, = 40$\,\gev\ \cite{Abelev:2014ffa,ALICE:2014dla}.

Simulations of \pp\ collisions were carried out using PYTHIA 6.425 and PYTHIA 8,
with the Perugia 2011 and 4C tunes
\cite{Skands:2010ak}. They were used as particle-level references to our fully corrected jet shapes. Moreover, instrumental effects were simulated using PYTHIA Perugia 0 for the primary collision followed by a detailed particle transport and detector response simulation using
GEANT3 \cite{GEANT3}. 
Simulated events, which include primary particles and the
daughters of strong and electromagnetic decays but not secondaries
from interactions in the detector material or the daughters of weak
decays from strange hadrons, are denoted as ``particle level''. Simulated events, which also
include instrumental effects and weak decay daughters, where
reconstructed tracks are selected using the experimental cuts, are denoted as ``detector level''.

\section{Jet reconstruction}
\label{sect:JetReco}

Jet reconstruction for both the \pp\ and Pb--Pb analysis was carried
out using the  \kT\ and anti-\kT\ \cite{FastJetAntikt} algorithms applied to all accepted
tracks. The $E$-scheme for recombination was used~\cite{Cacciari:2011ma} and the mass of the charged pion was assumed for each track.

The jet area, $A_{\rm jet}$, was calculated by the FastJet
algorithm~\cite{Cacciari:2011ma} using ghost particles (nearly zero-$p_{\rm T}$ particles that participate in the clustering procedure but do not modify the jet momentum) with area
$A_{\rm g} = 0.005$~\cite{FastJetArea}. A cut on the jet area was applied to suppress combinatorial jets while
preserving high efficiency for true hard jets
\cite{deBarros:2012ws, Jacobs:2010wq}. Jet candidates were rejected if
$A_{\rm jet}<0.07$ for $R = 0.2$.
Jet candidates were accepted if fully contained in the acceptance, meaning that their centroids laid within $|\eta_{\rm{jet}}|<0.7$, where $\eta_{\rm{jet}}$ is the pseudo-rapidity of the jet axis. 

In pp collisions and for the considered $R=0.2$, the change of the jet momentum due to the underlying event background is negligible. For the Pb--Pb analysis, corrections of jet $p_{\rm T}$ and jet shapes are needed due to the presence of large background from the underlying event. 
For that purpose, the jet reconstruction was carried out twice for each event. The
first pass applies the \kT\ algorithm with $R = 0.2$ to estimate
 the density of jet-like transverse-momentum and mass due to the
 background, $\rho$ and $\rho_{\rm m}$, respectively, defined as:

\begin{equation}
\rho=\mathrm{median}\left( \frac{p_{\rm {T,jet}}^{\rm{raw},\it{i}}}{A_{\rm{jet}}^{\it{i}}}\right),\, \, \, \, \,\rho_{m}=\mathrm{median}\left( \frac{\it{m}_{\it{i}}}{A_{\rm jet}^{\it{i}}}\right)
\label{eq:rho}
\end{equation}

\noindent
where the index $i$ runs over all jet candidates in an event and $p_{\rm{T,jet}}^{{\rm raw,}i}$, $m_{i}$,
and $A_{\rm jet}^{i}$ are the transverse momentum, mass, and area of the
$i^{\textup{th}}$ reconstructed jet. The two jets with highest $p_{\rm{T,jet}}^{\rm{raw},i}$ were excluded from the
calculation of the median to suppress the impact of signal jets on the underlying event background estimate. 
The second pass, which generates jet candidates
for the reported distributions, applied the \antikT\
algorithm with resolution parameter $R = 0.2$.

\section{Average background subtraction and fake jet suppression}
\label{sect:subtraction}

In order to correct the candidate jet $p_{\rm T}$ and shape distributions simultaneously for the average underlying event
background, two different techniques were applied.

\begin{itemize}
\item Area-derivatives method~\cite{Soyez:2012hv}.
This method is valid for any infrared and collinear safe jet reconstruction algorithm and
jet shape. The event is characterised by $\rho$ and $\rho_{\rm m}$. Ghost particles are added uniformly in the acceptance, each of them mimicking a pileup-like component in a region of area $A_{\rm g}$. The sensitivity of the shape to background is determined by calculating its derivatives with respect to the
transverse momentum and mass of the ghosts. Given $\rho$, $\rho_{\rm m}$, and the information on the derivatives, the value of the jet shape is then extrapolated by a Taylor series to zero background.

\item Constituent subtraction method~\cite{Berta:2014eza}.
In this method, the subtraction operates particle-by-particle. Ghost particles are added
uniformly to the acceptance, with finite $p_{\rm T}$ and mass given by:
$p_{\rm{T,g}}=A_{\rm{g}}\rho$ and 
$m_{\rm{g}}=A_{\rm{g}}\rho_{\rm m}$.  The distance between each 
 real jet constituent and each ghost is then computed and an iterative
 procedure starts, which consists of finding the closest pair. If the
 transverse momentum of particle $i$ is larger than that of the ghost,
 the ghost is discarded and its transverse momentum is subtracted from
 that of the real particle. In case the transverse momentum of the
 ghost is larger than that of particle $i$, the real particle is
 discarded and the transverse momentum of the real particle is subtracted from the ghost transverse momentum. The same
 procedure applies to the mass. The procedure terminates when all jet constituents are analysed.

 \end{itemize}

We note that in the case of $\rho_{\rm m}=0$, the jet $p_{\rm T}$ correction obtained with these methods coincides exactly with the standard area-based subtraction approach where $p_{\rm{T,jet}}^{sub}=p_{\rm{T,jet}}-\rho \times A_{\rm{jet}}$.
The $\rho_{\rm m}$ term  was introduced to take into account that low-$p_{\rm{T}}$ particles from the underlying event have masses that are not negligible compared to their momenta. This component has impact on the observables related to differences between jet energy and 3-momentum like the jet mass \cite{Soyez:2012hv} but negligible impact on the jet momentum. 

\begin{figure}[h]
  \includegraphics[width=0.33\textwidth]{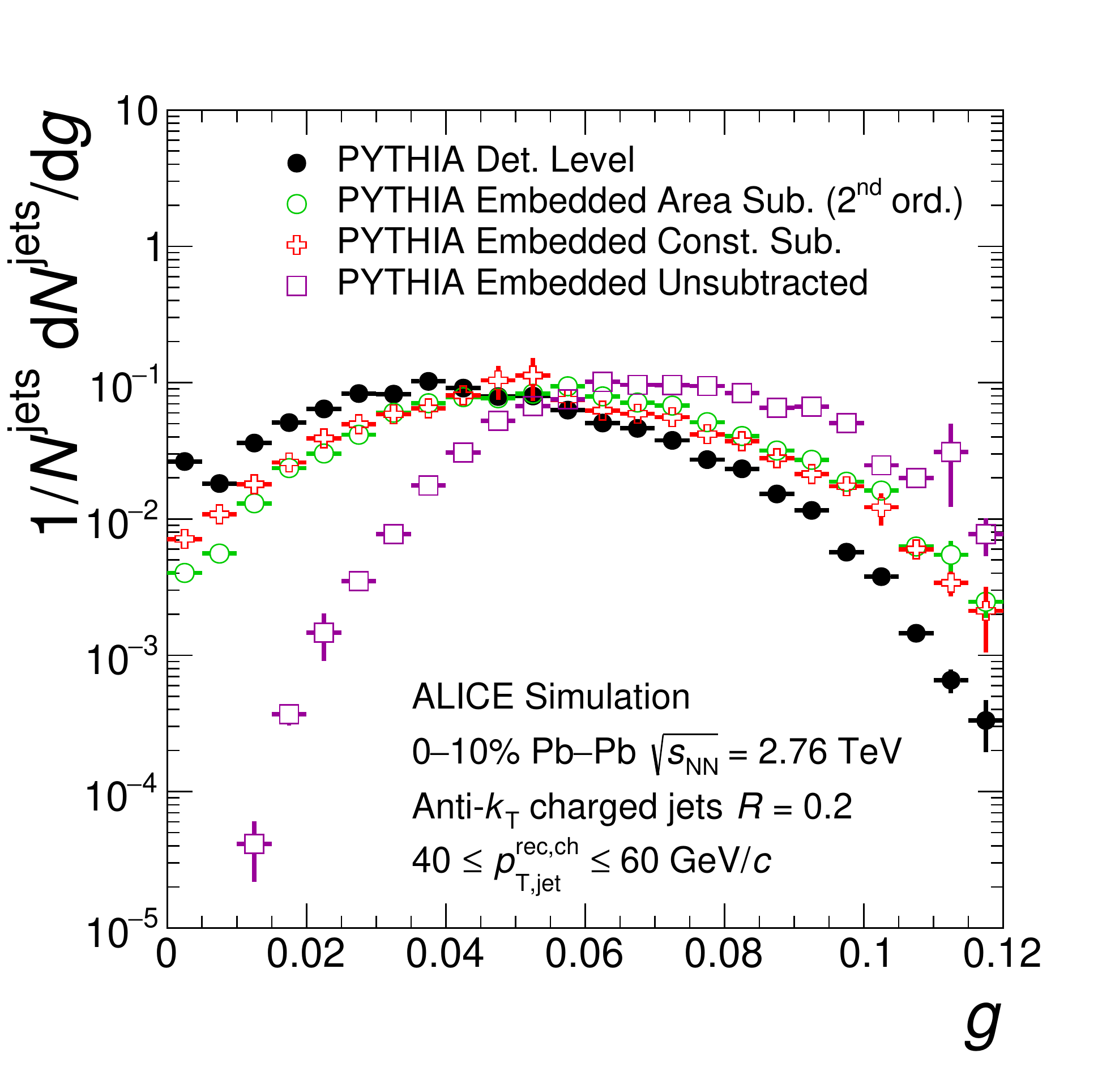}
\includegraphics[width=0.33\textwidth]{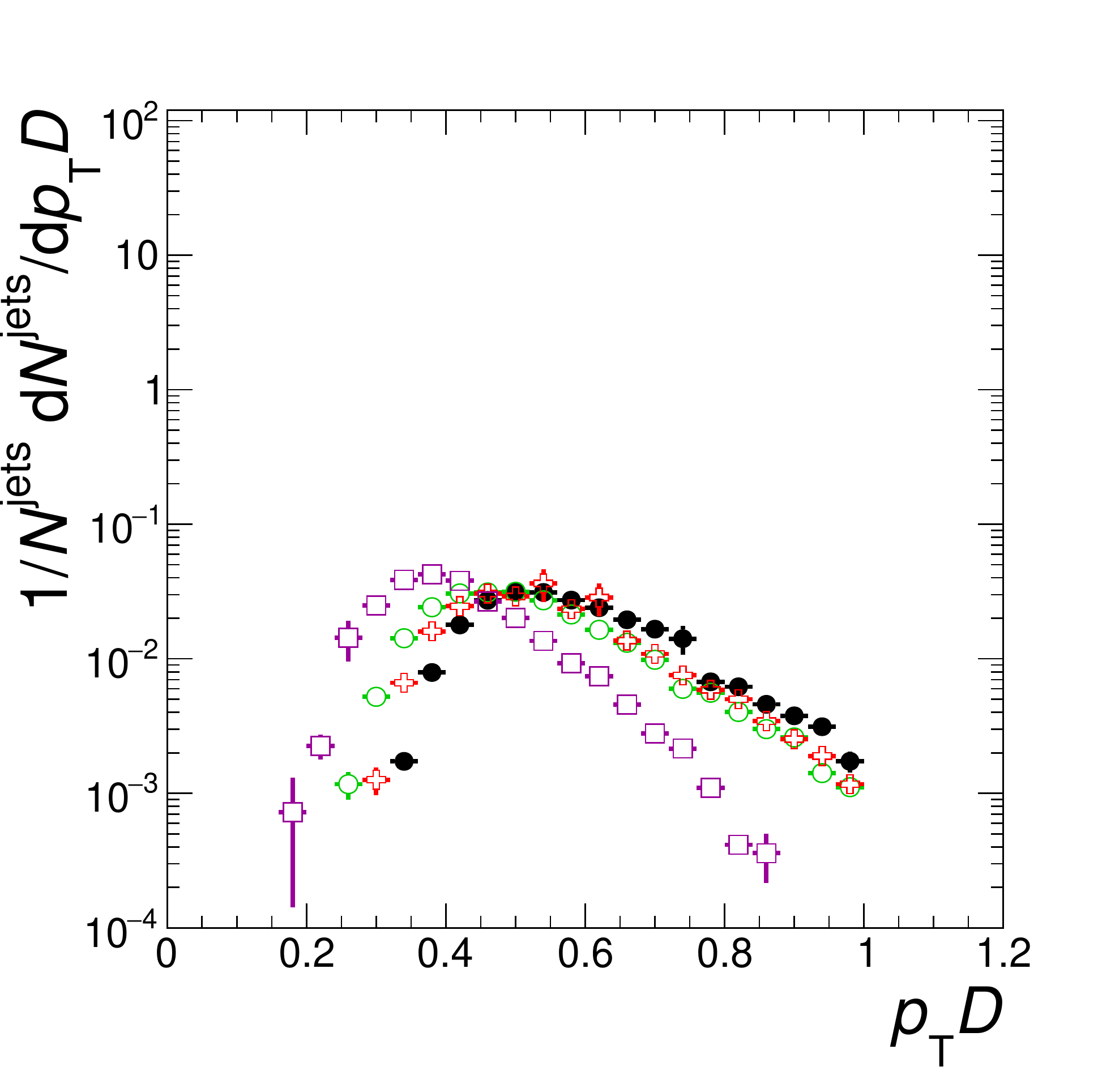}
\includegraphics[width=0.33\textwidth]{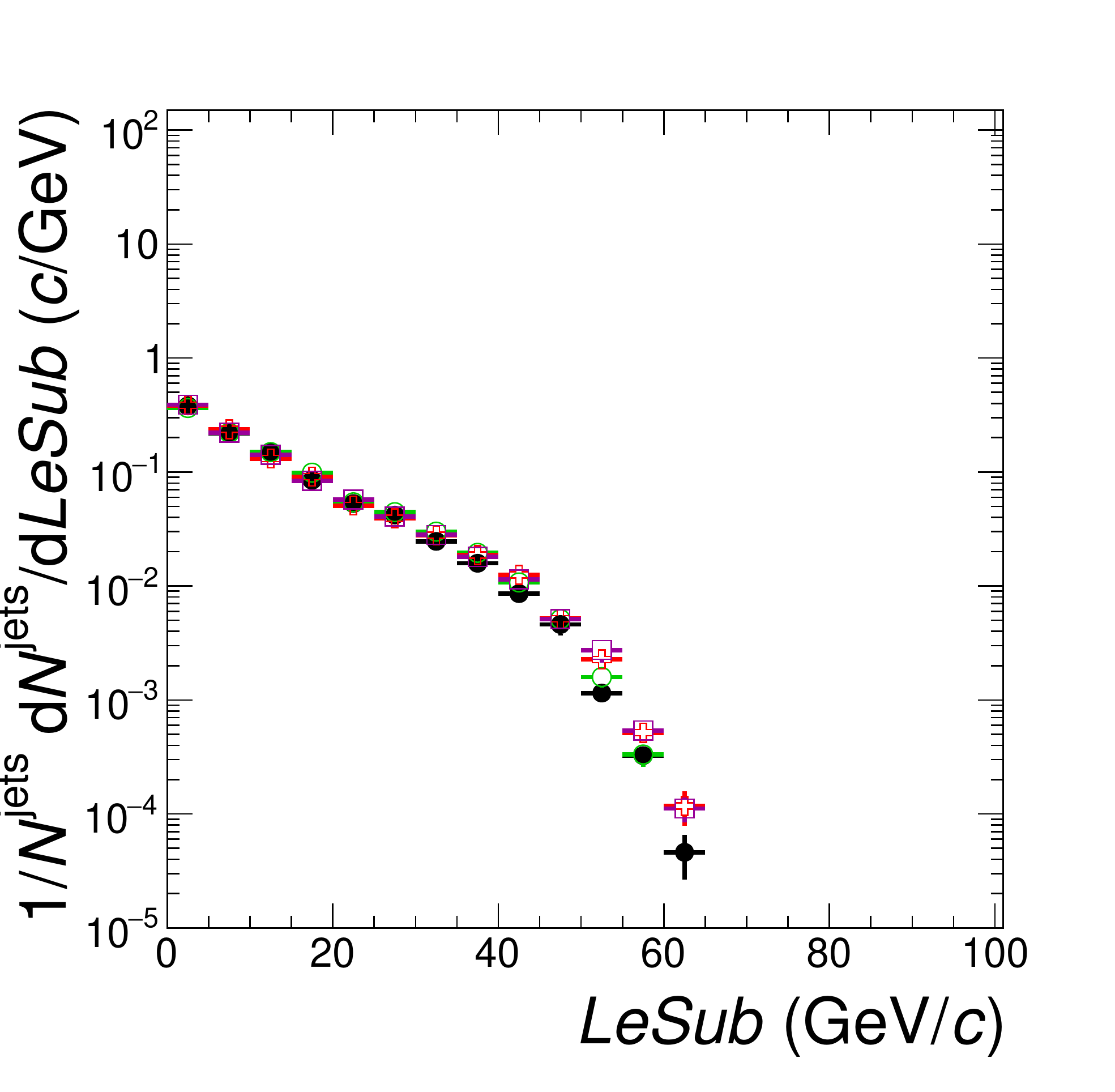}

\caption{Background subtraction performance for jet shapes studied with jets from PYTHIA events embedded into real Pb--Pb events, in the background subtracted transverse momentum interval $40 \leq
 p_{\mathrm{T,jet}}^{\rm{rec,ch}} \leq 60$\,GeV/$c$ for the area derivatives and constituent subtraction methods.}
\label{fig:BackSubPerform}
\end{figure}

The jet-by-jet constituent subtraction technique~\cite{Berta:2014eza} was used as the primary method and the area-derivatives method was used as a systematic variation. 
To study the performance of the subtraction methods, PYTHIA events at
detector level were embedded into Pb--Pb events. Embedding means superimposing the PYTHIA and Pb--Pb events at track level.  
Figure~\ref{fig:BackSubPerform} shows
the shape distribution for embedded unsubtracted jets (squares), the average background-subtracted jet shapes (open circles and crosses for the two methods), and the 
PYTHIA detector-level distributions (full circles).
The average background-subtracted embedded distributions get closer to the PYTHIA detector-level distributions than without background subtraction. The comparison was performed in the interval of reconstructed and subtracted embedded momentum, $p_{\rm{T,Jet}}^{\rm{rec,ch}}$, of $40$--$60$\,GeV/$c$. Figure~\ref{fig:BackSubPerform} reveals that $LeSub$ is rather
insensitive to modifications induced by the background.    
Residual differences, due to background fluctuations, were
corrected using an unfolding procedure (see Section~\ref{sect:unfolding}). 

The smearing of jet $p_{\rm T}$ due to the local background fluctuations, quantified as $\delta p_{\mathrm{T}}$ \cite{Abelev:2012ej}, has a width of $\sigma=$ 4\,GeV/$c$ for
$R = 0.2$ in central Pb--Pb collisions \cite{Abelev:2012ej}. The truncation of the raw yield at 30\,GeV/$c$
prior to unfolding sets our working point more than 7$\sigma$ away from zero and thus contributions from purely combinatorial background jets to the raw spectrum are negligible, allowing for a stable unfolding correction~\cite{Abelev:2013kqa}.

\section{Detector and background response}
\label{sect:response}
A 4-dimensional response matrix was built with axes
$shape^{\mathrm{part,ch}}$, $p_{\mathrm{T,jet}}^{\mathrm{part,ch}}$, $shape^{\mathrm{rec,ch}}$, and $p_{\mathrm{T,jet}}^{\mathrm{rec,ch}}$.
The upper index 'part' refers to particle level and 'rec' refers to
reconstructed level quantities. In pp collisions, index 'rec' refers to detector-level quantities, while in Pb--Pb collisions it refers to embedded and background subtracted quantities. In order to associate a reconstructed-level jet to a particle-level jet, a matching criterion needs to be defined. 

The response matrix for pp collisions is purely instrumental and was constructed using PYTHIA events at particle level and after full detector simulation. The matching criterion between the corresponding jets at particle and detector level is purely geometrical and was based on requiring that they are univocally the closest in the ($\eta,\varphi$) space. 
The response matrix for the Pb--Pb case considers both the effects of the detector and the effects of the background fluctuations. To construct it, we embedded PYTHIA detector-level events into Pb--Pb events and
we applied two successive matchings, the first between the background-subtracted, embedded jets and detector-level PYTHIA jets
and the second between the detector and particle-level jets. The matching between embedded and detector-level jets is not purely geometrical but also required that more than  $50\%$ of the detector-level jet momentum is contained in the embedded reconstructed jet. The matching efficiency is consistent with unity for jets with $p_{\rm T}$ above 30\,GeV/$c$. 
We note that since our embedding is a superposition of PYTHIA and Pb--Pb events at track level, two-track effects are not present, however their impact in data is small due to the large required number of clusters per track.
The jet energy scale shift in pp collisions is about 15$\%$ at $p_{\rm{T,jet}}^{\rm{part,ch}}=50$\,GeV/$c$. In Pb--Pb collisions, this shift is about $12\%$ in the same transverse momentum range at the particle level~\cite{Abelev:2013kqa}.
The instrumental jet energy resolution (JER), which characterises the detector response relative to charged jets at particle level, varies from $20\%$ at $p_{\mathrm{T,jet}}^{\rm{rec,ch}}
= 20$\,GeV/$c$ to $25\%$ at $p_{\mathrm{T,jet}}^{\rm{rec,ch}}= 100$\,GeV/$c$, similarly for pp and Pb--Pb collisions. The JER is dominated by tracking effects and shows no dependence with jet $R$ \cite{Abelev:2013kqa}.

\begin{figure}[h]
  \centering
  \includegraphics[width=0.33\textwidth]{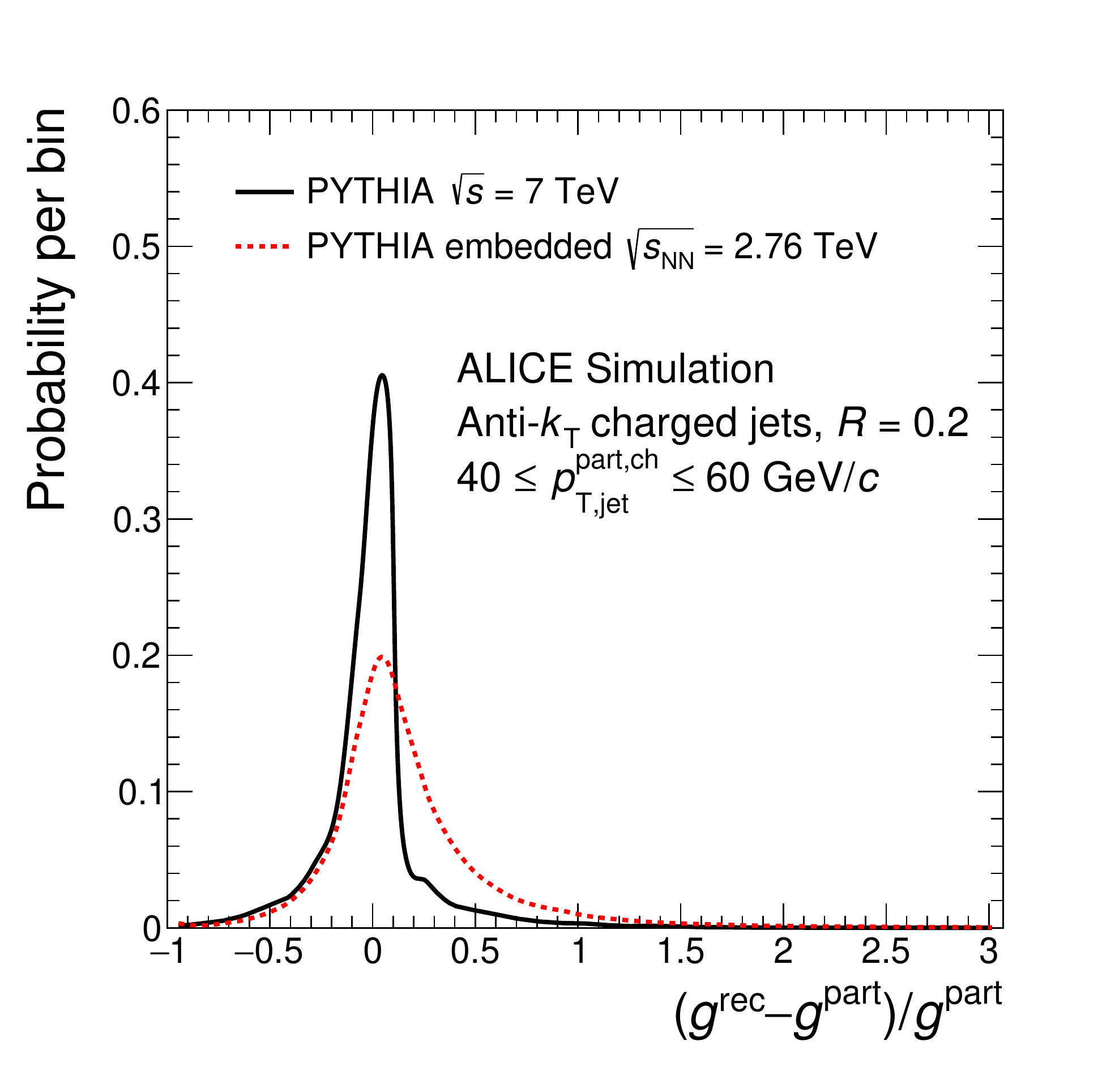}
\includegraphics[width=0.33\textwidth]{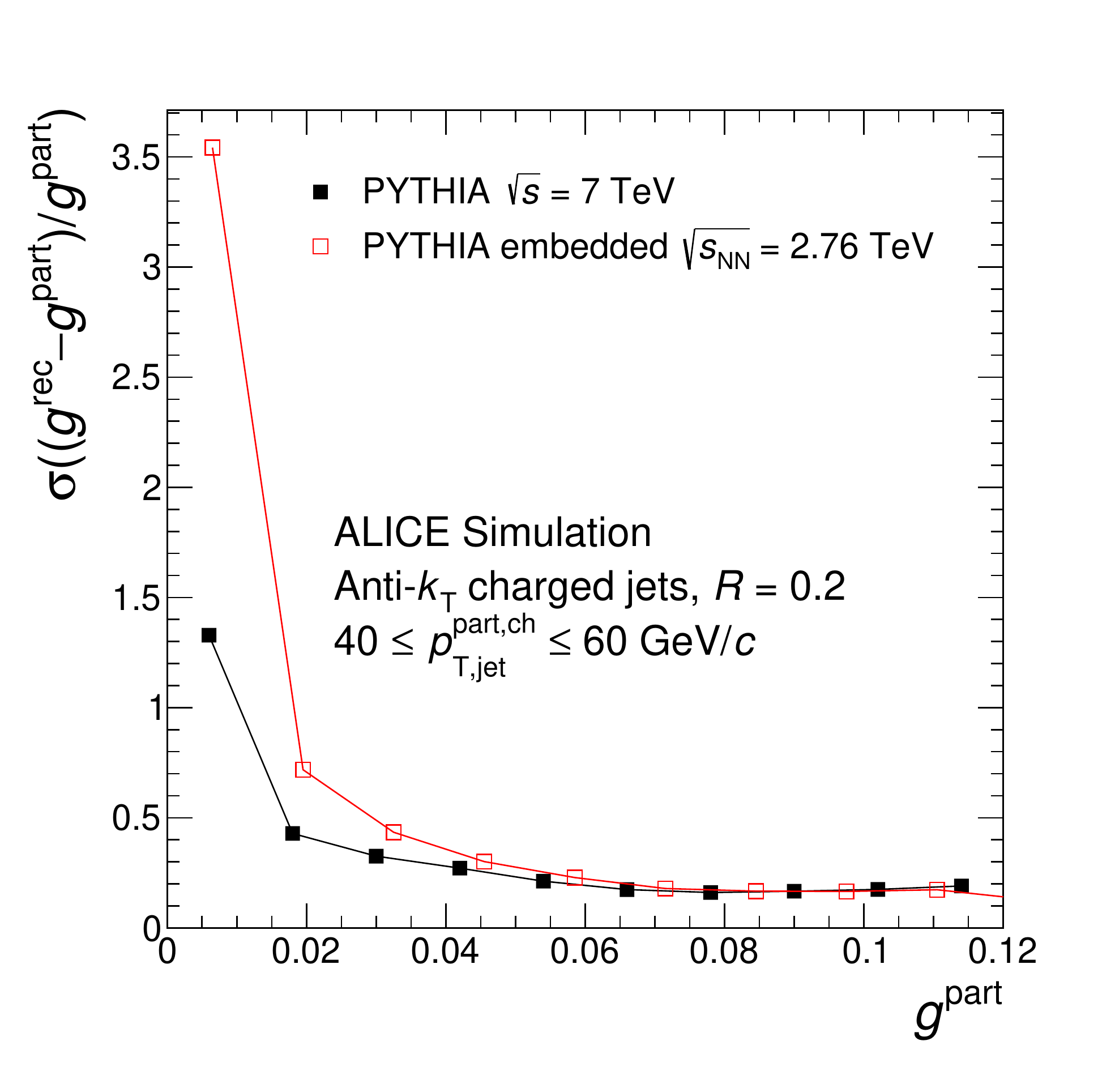}
\includegraphics[width=0.33\textwidth]{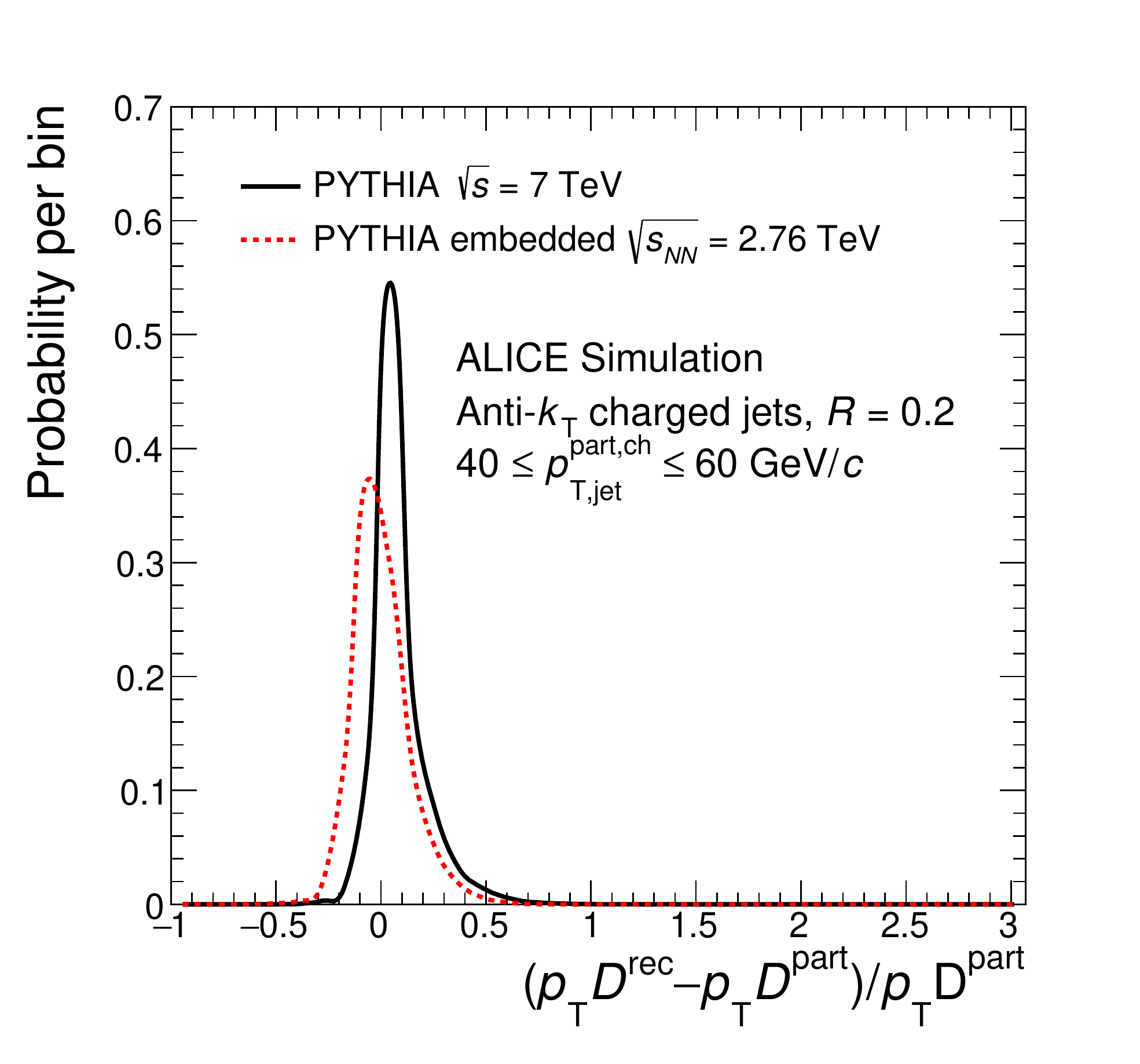}
\includegraphics[width=0.33\textwidth]{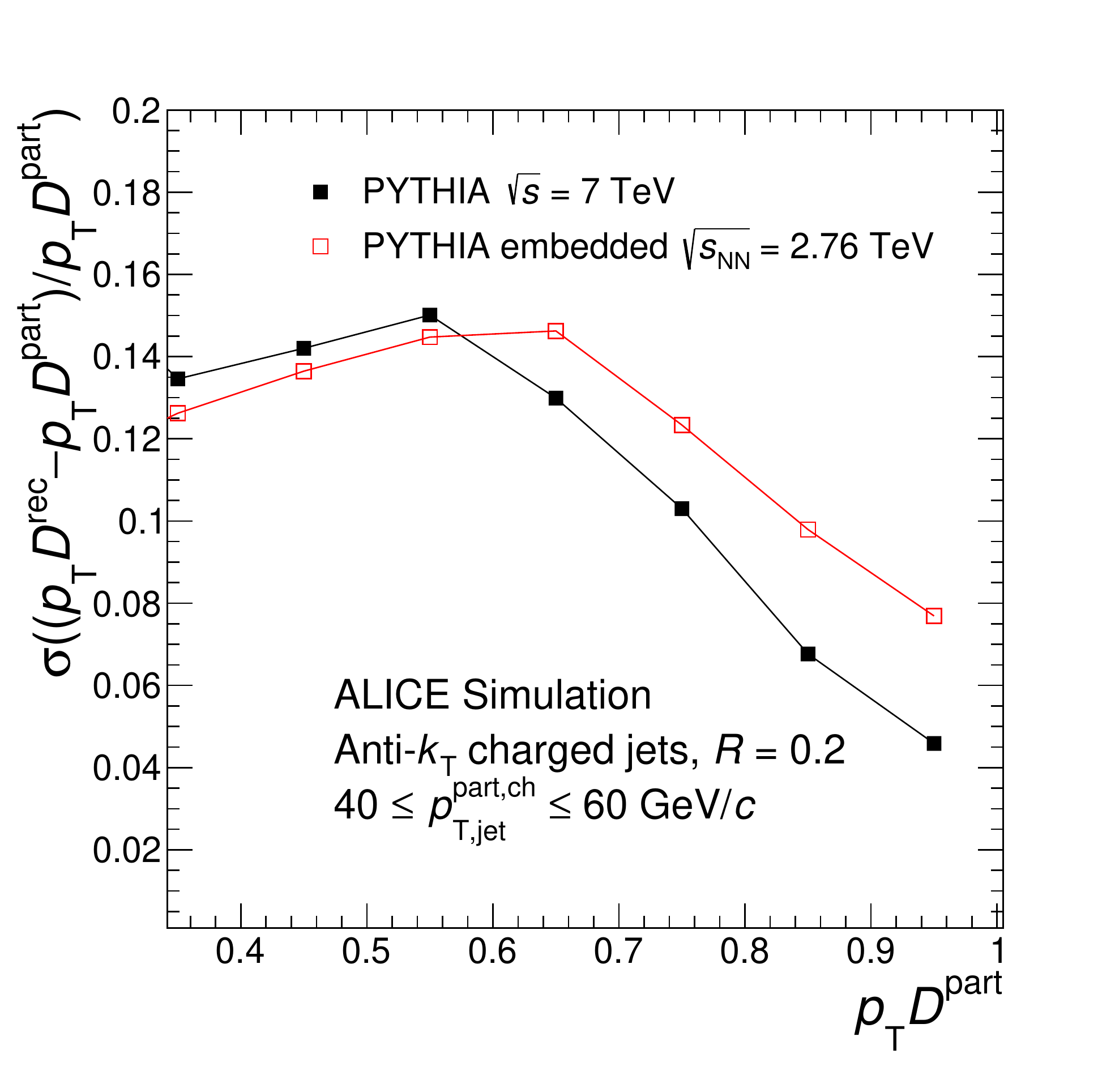}
\includegraphics[width=0.33\textwidth]{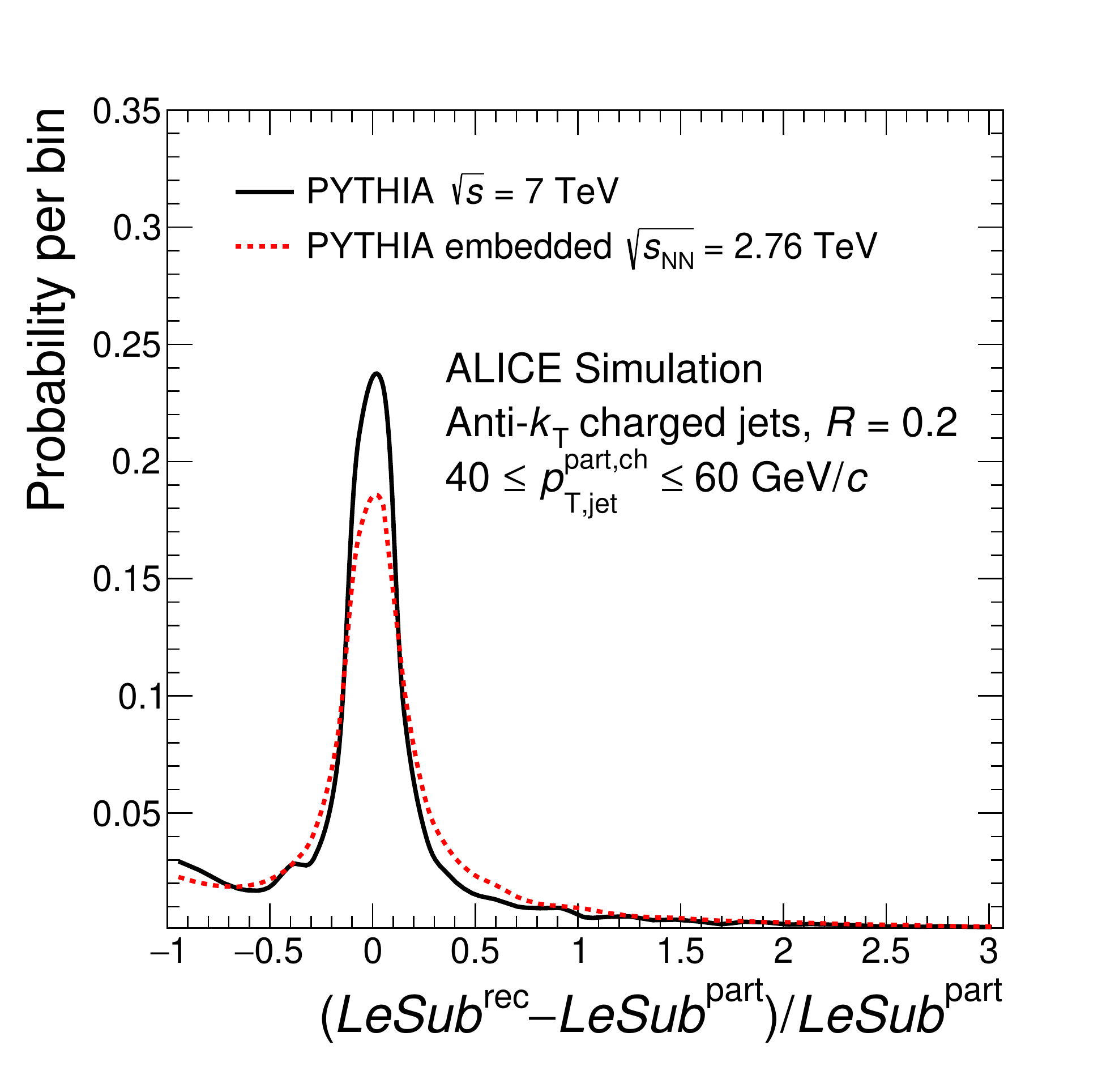}
\includegraphics[width=0.33\textwidth]{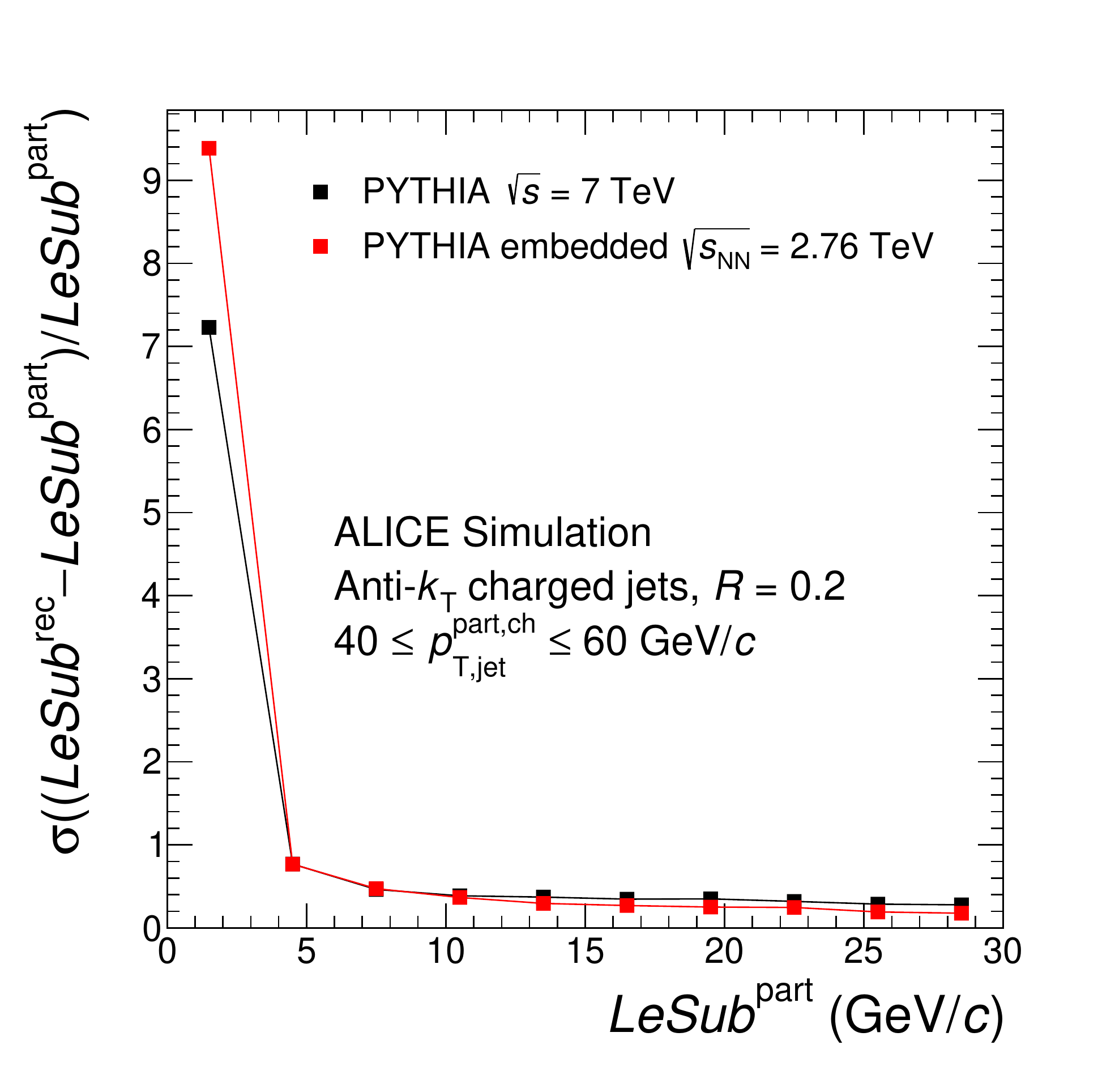}
\caption{Left plots show the distributions of residuals for the set of
  three jet shapes in a given interval of $p_{\rm{T,jet}}^{\rm{part,ch}}$
  $40$--$60$ GeV/$c$ using the PYTHIA and PYTHIA embedded simulations. Right plots show the width (quantified as the standard deviation) of the
  distributions on the left as a function of the values of the shapes
  at particle level. The black and red curves
correspond to pp and Pb--Pb simulations, respectively. The line connecting the points on the right is drawn to guide the eye.}  
\label{fig:Residuals}
\end{figure}

The jet shape resolution can be
studied via the distribution of residuals, which gives the relative difference between the jet shapes measured at particle
and reconstructed level.
In Fig.~\ref{fig:Residuals}, the left panels show the distribution of residuals
for each of the three shapes for jets in pp and for PYTHIA embedded jets in Pb--Pb collisions, in
the particle-level jet $p_{\rm{T,jet}}^{\rm{part,ch}}$ range of $40$--$60$~GeV/$c$. Tracking inefficiency
induces a negative tail in the angularity (narrower jets due to missing constituents), while a positive
tail is induced on average by background fluctuations and, to a lower extend, also by track momentum resolution. The trend is the opposite in the case of $p_{\rm T} D$: losses due to tracking efficiency shift the distribution of residuals to positive values (fewer constituents) while the background fluctuations induce a negative shift.
For $LeSub$, the distributions in PYTHIA and PYTHIA embedded simulations are similar due to the resilience of the observable to background fluctuations.  

The right panels of Fig.~\ref{fig:Residuals} show the resolution of the shapes,
quantified as the standard deviation $\sigma$ of the distribution of residuals, as a
function of the shape at particle level for pp and Pb--Pb collisions. 
At low angularity, the resolution is poor because these jets are more
collimated and typically have fewer constituents. In this region, this
shape is thus more
sensitive to tracking inefficiency and background fluctuations.
At higher angularities the resolution improves up to 20$\%$.
The resolution of $p_{\rm T}D$ is overall below $15\%$ and improves for harder jets when $p_{\mathrm{T}}D$ approaches unity.
 A similar case holds for $LeSub$, for which the resolution improves
 at higher values of the shape and worsens at low values where
 detector effects have a larger impact.

\section{Two-dimensional unfolding procedure}
\label{sect:unfolding}
\noindent Residual background fluctuations and detector effects were
unfolded to simultaneously correct the reconstructed jet transverse momentum and shape distributions back to the particle level.
Bayesian unfolding in two dimensions as implemented in the  RooUnfold
package~\cite{RooUnfold} was used.
Several considerations needed to be taken into account. The 2D
correlation ($p_{\rm{T,jet}}^{\rm{rec,ch}}$, $shape_{\rm
  jet}^{\rm{rec,ch}}$), which is the input to the unfolding, was binned
such that there are at least 10 counts per bin, to guarantee statistical stability of the correction procedure, which also sets the upper limit of the input $p_{\rm{T,jet}}^{\rm{rec,ch}}$ range (80~GeV/$c$ both in pp and Pb--Pb collisions). The shape input ranges are $0.3$--$1$, $0.02$--$0.12$, and $0$--$30$\,GeV/$c$ for $g$, $p_{\rm T}D$ and $LeSub$, respectively, for both collision systems. 
The raw input correlation should not contain combinatorial background,
which was suppressed by truncating it at sufficiently
high values of $p_{\rm{T,jet}}^{\rm{rec,ch}}$. The lower limit of the input $p_{\rm{T,jet}}^{\rm{rec,ch}}$ range for unfolding in Pb--Pb collisions is $30$\,GeV/$c$. As argued in Section~\ref{sect:JetReco}, 
fake jet contamination above this limit for jets measured with $R = 0.2$
is negligible. In pp collisions, the cutoff is set at $p_{\rm{T,jet}}^{\rm{rec,ch}}=20$\,GeV/$c$.

 The particle-level $p_{\rm T,jet}^{\rm part,ch}$ range of the response matrix is from 0 to 200\,GeV/$c$. The shape ranges at the particle level are $0$--$1$, $0$--$0.12$ and $0$--$50$\,GeV/$c$ for $g$, $p_{\rm T}D$, and $LeSub$, respectively.
The particle-level ranges were extended beyond the input ranges to allow for jet migration into the reconstructed level range due to
background fluctuations and tracking efficiency losses.  
When the data input is truncated, feed-in from detector-level jets outside
the truncated range had to be considered and corrected for. However,
this correction (referred to as kinematic efficiency) is purely based on MC and has to be limited by
considering unfolded bins far away from the truncation
thresholds. Thus, our final results are presented for
the jet momentum interval $40$--$60$\,GeV/$c$.

We tested the stability of the unfolding by refolding the solution back and checking the agreement with the raw distribution. In pp (Pb--Pb) collisions, both distributions agree within 1$\%$ (5$\%$) after the second (third) iteration.
The unfolding solutions converged after few iterations (note that
convergence occurs globally in 2D and not just in a given interval of jet
$p_{\rm T}$). We also performed a closure test, where two
statistically independent MC samples are used to fill the response and
the pseudo-data. In this case, the unfolded solution agrees with the
MC truth within less than 10$\%$ in pp and Pb--Pb collisions.

\section{Systematic uncertainties}
\label{sect:sysuncert}
The systematic uncertainties for the shapes were determined by varying the analysis settings for instrumental response and background fluctuations. The systematic uncertainties are listed below: 
\begin{itemize}

\item Tracking efficiency uncertainty for the used track selection is $\pm 4\%$ and this was used as source to estimate the jet energy scale uncertainty \cite{Acharya:2017goa}.

\item The prior in the 2D Bayesian implementation of RooUnfold was
  taken as the projection of the response matrix onto the true
  axes. The default prior was PYTHIA Perugia 0. We considered three
  different variations. As the systematic uncertainty in a given bin
  we take the maximal deviation out of the three variations. The first
  variation was to re-weight the response matrix such that the prior
  coincides with the unfolding solution. The second considered
  variation was obtained by re-weighting the response matrix such that
  the projection onto true axis was that of purely gluonic jets. The
  third variation was obtained by re-weighting the response matrix
  such that the projection onto the true axis was that of purely quark jets.

\item  The regularization was given by the number of iterations
  considered, which was 4 (8) for pp (Pb--Pb) collisions in the
  default solutions. The uncertainty in the regularization was estimated by considering differences to solutions for one less and two more iterations.

\item The minimum accepted jet
  $p_{\mathrm{T,jet}}^{\rm ch}$ as input to the unfolding was 20
  (30)\,GeV/$c$ in pp (Pb--Pb) collisions. As a variation, we lowered the truncation
  by 10\,GeV/$c$.

\item The binning of the raw input was changed arbitrarily (but keeping the statistical requirements of at least 10 counts per bin) in both the
  $p_{\rm{ T,jet}}^{\rm ch}$ and shape dimensions.

\item The choice of the background subtraction method in Pb--Pb collisions affected mostly the tails of the distribution and resulted in a variation of 10$\%$ at most. 

\item In Pb--Pb collisions, the matching criterion in the tagging algorithm was relaxed so that the response was filled with pairs of jets where the reconstructed embedded jet contained at least $40\%$ of the probe jet momentum.

\end{itemize}
The different components of the systematic uncertainties
for the different shapes are summarized in
Tables~\ref{tab:Systpp}~and~\ref{tab:SystPbPb} for pp and Pb--Pb collisions,
respectively.  The largest contribution to the systematic uncertainties on the fully corrected pp data comes from the tracking
efficiency uncertainty, yet the total systematic uncertainty is smaller when compared to the statistical one. In Pb--Pb collisions, systematic uncertainties due to prior and subtraction method choice dominate over statistical uncertainties. All the uncertainties induce changes in the shape of our observables and the applied normalization causes long range anti-correlations. The total uncertainty is obtained by adding the different components in quadrature. 

\begin{table}[!t]
\renewcommand{\arraystretch}{1.3}
\begin{tabular}{|l|ccc|ccc|ccc|}
\hline 
\multicolumn{1}{|l|}{Shape} & \multicolumn{3}{c|}{$p_{\mathrm{T}}D$} 
 & \multicolumn{3}{c|}{$g$}& \multicolumn{3}{c|}{$LeSub$ (GeV/$c$)} \\
\hline
\hline
\multicolumn{1}{|l|}{Shape interval} 
& 0.3-0.4 & 0.5-0.6 & 0.8-1 & 0-0.02 & 0.05-0.06 & 0.08-0.12 & 0-5 & 10-15 & 20-30 \\
\cline{1-10}

Tracking              & $10\%$ & $0.70\%$  & $11\%$ 
                      & $10\%$ & $1.7\%$ & $4.2\%$ 
                      & $1.8\%$ &$0.5\%$  & $6.6\%$  \\

Prior                  & $_{-0.0}^{+0.3}\%$ & $_{-0.0}^{+0.9}\%$  & $_{-0.0}^{+0.0}\%$ 
                      & $_{-3.0}^{+0.0}\%$ & $_{-1.2}^{+0.0}\%$ & $_{-0.0}^{+3.0}\%$ 
                      & $_{-0.0}^{+0.9}\%$ &
                                                               $_{-0.0}^{+0.6}\%$  & $_{-0.0}^{+0.5}\%$  \\

Regularization  & $_{-0.3}^{+0.1}\%$ & $_{-1.2}^{+0.7}\%$  & $_{-0.1}^{+0.4}\%$ 
                      & $_{-2.7}^{+5.9}\%$ & $_{-1.0}^{+2.3}\%$ & $_{-4.5}^{+2.6}\%$ 
                      & $_{-1.3}^{+0.8}\%$ &
                                                               $_{-0.6}^{+0.6}\%$  & $_{-0.0}^{+0.6}\%$  \\

 Truncation  & $_{-0.7}^{+0.0}\%$ & $_{-0.1}^{+0.0}\%$  & $_{-0.0}^{+0.5}\%$ 
                      & $_{-0.0}^{+0.3}\%$ & $_{-0.2}^{+0.0}\%$ & $_{-0.0}^{+0.3}\%$ 
                      & $_{-0.0}^{+0.1}\%$ &
                                                               $_{-0.1}^{+0.0}\%$  & $_{-0.0}^{+0.1}\%$  \\

 Binning  & $1.4\%$ & $1.6\%$  & $4.2\%$ 
                      & $0.2\%$ & $6.4\%$ & $2.5\%$ 
                      & $2.1\%$ &
                                                               $1.8\%$  & $0.9\%$  \\
 \cline{1-10}
 Total  & $_{-10}^{+10}\%$ & $_{-2.2}^{+2.1}\%$  & $_{-11}^{+11}\%$
                       & $_{-11}^{+12}\%$ & $_{-6.8}^{+7.0}\%$  & $_{-6.7}^{+6.3}\%$  
                      & $_{-3.1}^{+3.0}\%$ & $_{-2.0}^{+2.1}\%$ & $_{-6.6}^{+6.7}\%$ \\
                     
\hline
\end{tabular}
\caption{Relative systematic uncertainties on the measured jet shapes in pp collisions for three selected jet shape intervals in the jet $p_{\mathrm{T,jet}}^{\rm ch}$ range of $40$--$60$\,GeV/$c$.}
\label{tab:Systpp}
\end{table}

\begin{table}[!t]
\renewcommand{\arraystretch}{1.3}
\begin{tabular}{|l|ccc|ccc|ccc|}
\hline 
\multicolumn{1}{|l|}{Shape} & \multicolumn{3}{c|}{$p_{\mathrm{T}}D$} 
 & \multicolumn{3}{c|}{$g$}& \multicolumn{3}{c|}{$LeSub$ (GeV/$c$)} \\
\hline
\hline
\multicolumn{1}{|l|}{Shape interval} 
& 0.3-0.4 & 0.5-0.6 & 0.8-1 & 0-0.02 & 0.05-0.06 & 0.08-0.12 & 0-5 & 10-15 & 20-30\\
\cline{1-10}

Tracking   & $0.7\%$ & $1.1\%$  & $3.3\%$ 
                      & $9.6\%$ & $2.9\%$ & $4.9\%$ 
                      & $0.6\%$ &
                                                               $1.7\%$  & $0.8\%$  \\

Prior                  & $20\%$ & $2.6\%$  & $7.4\%$ 
                      & $7.6\%$ & $8.1\%$ & $20\%$ 
                      & $7.5\%$ &
                                                               $7.9\%$  & $9.0\%$  \\

Regularization  & $_{-1.5}^{+0.6}\%$ & $_{-0.8}^{+0.3}\%$  & $_{-0.3}^{+0.1}\%$ 
                      & $_{-0.9}^{+0.3}\%$ & $_{-0.8}^{+0.5}\%$ & $_{-0.0}^{+0.1}\%$ 
                      & $_{-1.1}^{+0.4}\%$ &
                                                               $_{-0.1}^{+0.2}\%$  & $_{-1.7}^{+4.3}\%$  \\

 Truncation  & $_{-18}^{+0.0}\%$ & $_{-0.0}^{+1.6}\%$  & $_{-0.0}^{+3.9}\%$ 
                      & $_{-0.0}^{+3.7}\%$ & $_{-1.0}^{+0.0}\%$ & $_{-39}^{+0.0}\%$ 
                      & $_{-25}^{+0.0}\%$ &
                      $_{-0.0}^{+10}\%$  & $_{-0.0}^{+18}\%$  \\

 Binning  & $1.3\%$ & $2.3\%$  & $4.2\%$ 
                      & $2.3\%$ & $3.6\%$ & $3.5\%$ 
                      & $0.9\%$ &
                                                               $7.9\%$
                                     & $3.4\%$  \\

Bkg.Sub                  & $_{-0.0}^{+5.5}\%$ & $_{-2.1}^{+0.0}\%$  & $_{-0.3}^{+0.0}\%$ 
                      & $_{-2.5}^{+0.0}\%$ & $_{-9.5}^{+0.0}\%$ & $_{-13}^{+0.0}\%$ 
                      & $_{-1.0}^{+0.0}\%$ &
                                                               $_{-6.7}^{+0.0}\%$  & $_{-1.6}^{+0.0}\%$  \\

Matching                   & $_{-0.5}^{+0.0}\%$ & $_{-0.0}^{+0.2}\%$  & $_{-0.0}^{+9.4}\%$ 
                      & $_{-0.0}^{+2.6}\%$ & $_{-0.0}^{+1.9}\%$ & $_{-0.0}^{+23}\%$ 
                      & $_{-4.3}^{+0.0}\%$ &
                                                               $_{-0.3}^{+0.0}\%$  & $_{-0.7}^{+0.0}\%$  \\
 \cline{1-10}
Total                   & $_{-27}^{+21}\%$ & $_{-4.3}^{+4.0}\%$  & $_{-9.2}^{+14}\%$ 
                      & $_{-13}^{+13}\%$ & $_{-13}^{+9.5}\%$ & $_{-47}^{+31}\%$ 
                      & $_{-26}^{+7.6}\%$ &
                                                               $_{-13}^{+15}\%$  & $_{-10}^{+21}\%$  \\
\hline
\end{tabular}
\caption{Relative systematic uncertainties on the measured jet shapes in Pb--Pb collisions for three selected jet shape intervals in the jet $p_{\rm{T,jet}}^{\rm ch}$ range $40$--$60$\,GeV/$c$.}
\label{tab:SystPbPb}
\end{table}

\section{Results and discussion}
\label{sect:results}

\begin{figure}[h]
\centering

\includegraphics[width=0.45\textwidth,height=0.45\textheight]{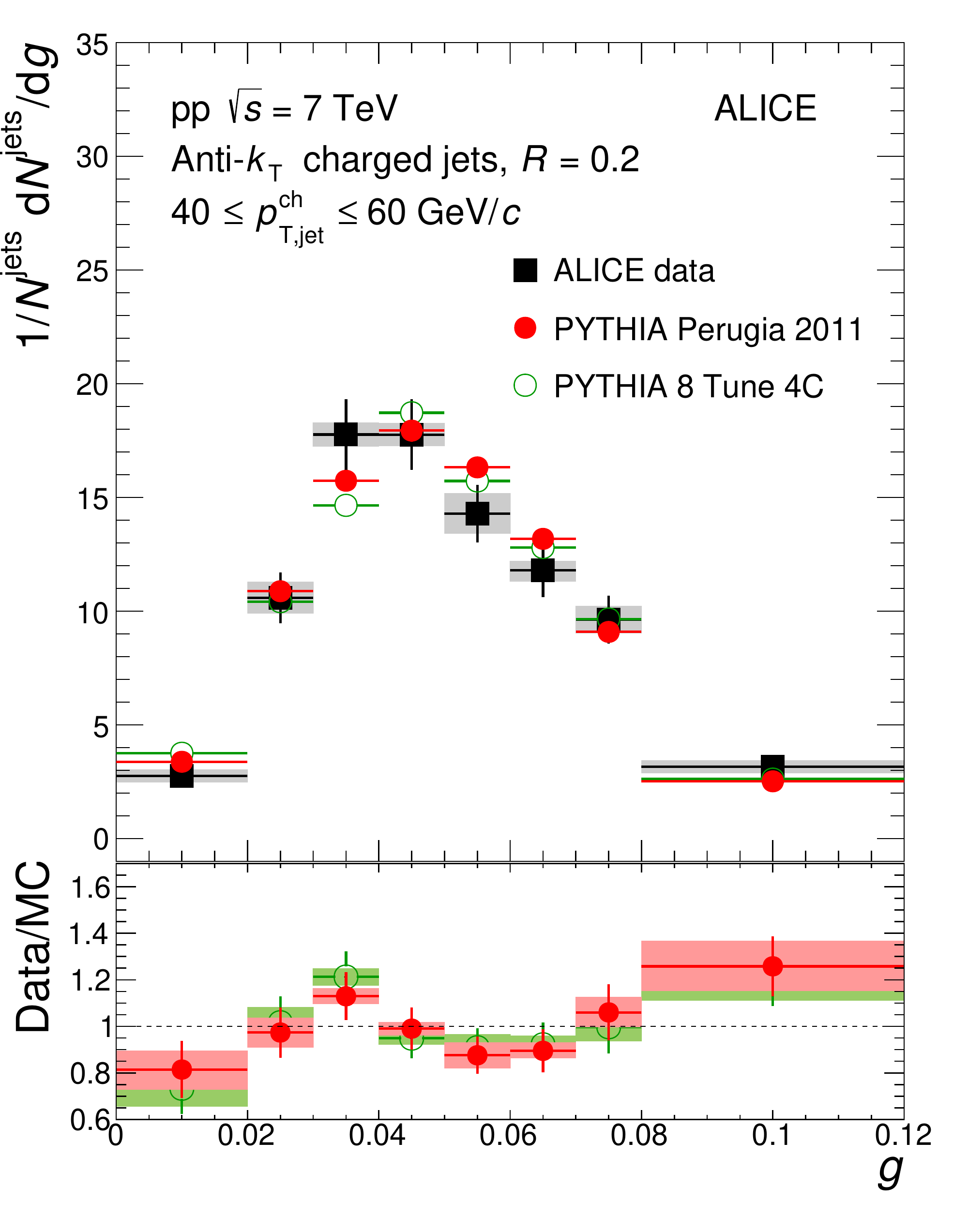}
\includegraphics[width=0.45\textwidth,height=0.45\textheight]{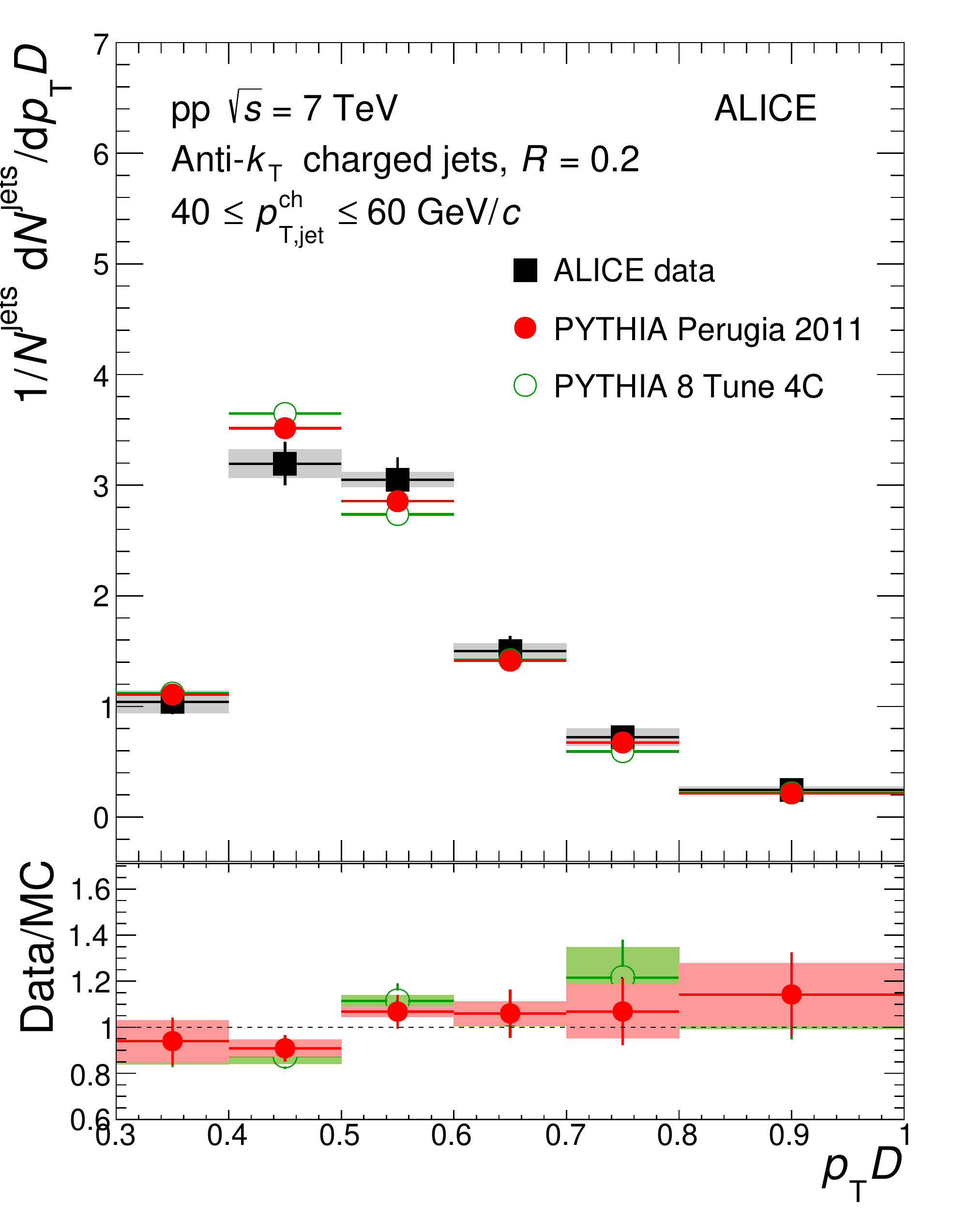}
\includegraphics[width=0.45\textwidth,height=0.45\textheight]{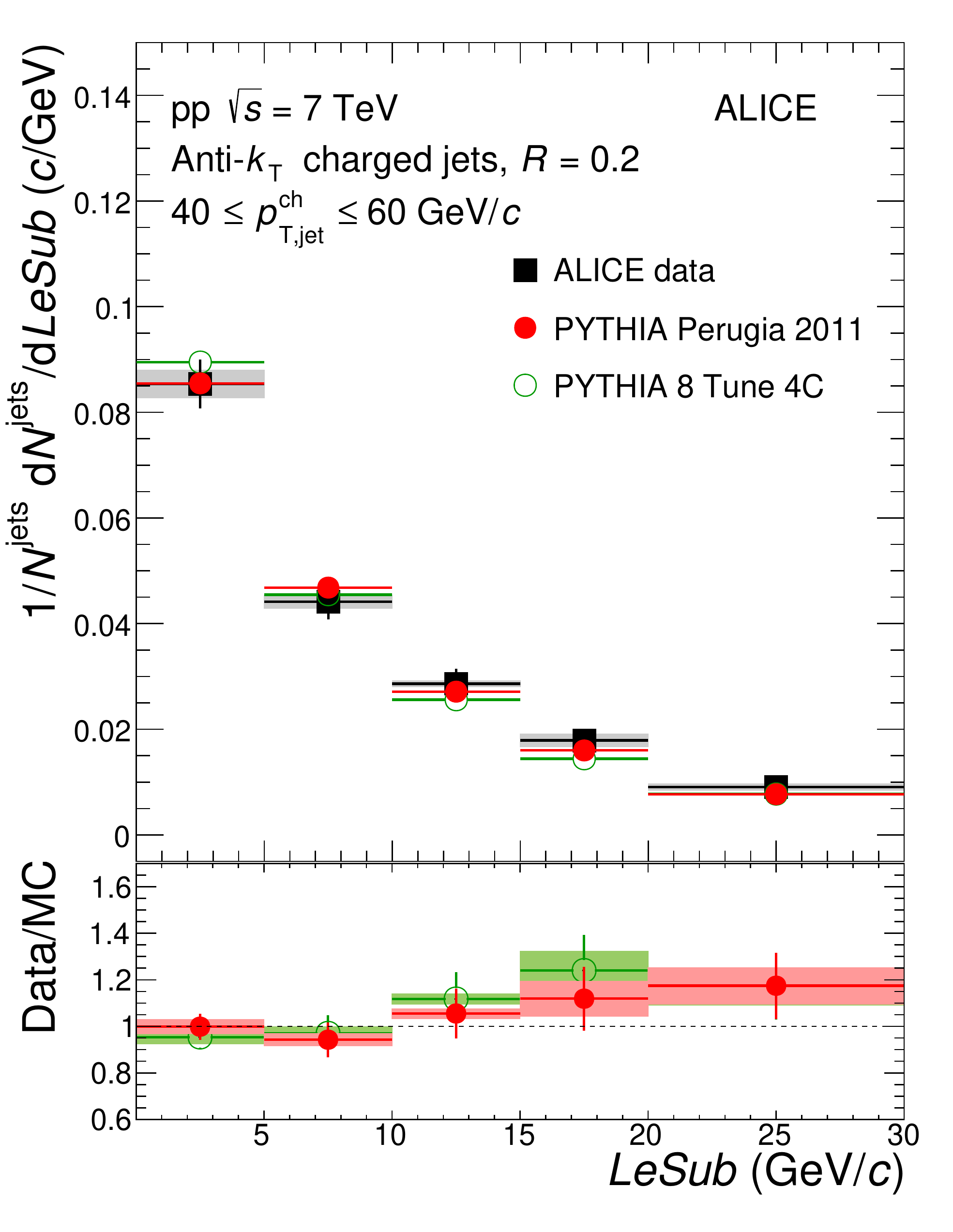}
\caption{Fully corrected jet shape distributions measured in pp collisions at $\sqrt{s}=7$\,TeV for $R = 0.2$ in 
the range of jet $p_{\mathrm{T,jet}}^{\rm ch}$ of $40$--$60$\,GeV$/c$. The results are compared to PYTHIA. The coloured boxes represent the uncertainty on the jet shape (upper panels) and its propagation to the ratio (lower panels)}
\label{fig:Resultspp}
\end{figure}

\noindent Figure~\ref{fig:Resultspp} presents the fully corrected jet shape distributions measured in
pp collisions at $\sqrt{s}=7$\,TeV in the jet $p_{\mathrm{T}}$ range $40$--$60$\,GeV/$c$.  The results are compared to PYTHIA Perugia 2011 and PYTHIA 8 tune 4C jet shape distributions obtained at the same collision energy. The ratio plots in the lower panels indicate a reasonable agreement within $20\%$. Large
non-perturbative effects are expected for small-$R$ jets~\cite{Dasgupta:2007wa} and seem to be well accounted for by the simulations. 

\begin{figure}[h]
\centering

\includegraphics[width=0.45\textwidth,height=0.45\textheight]{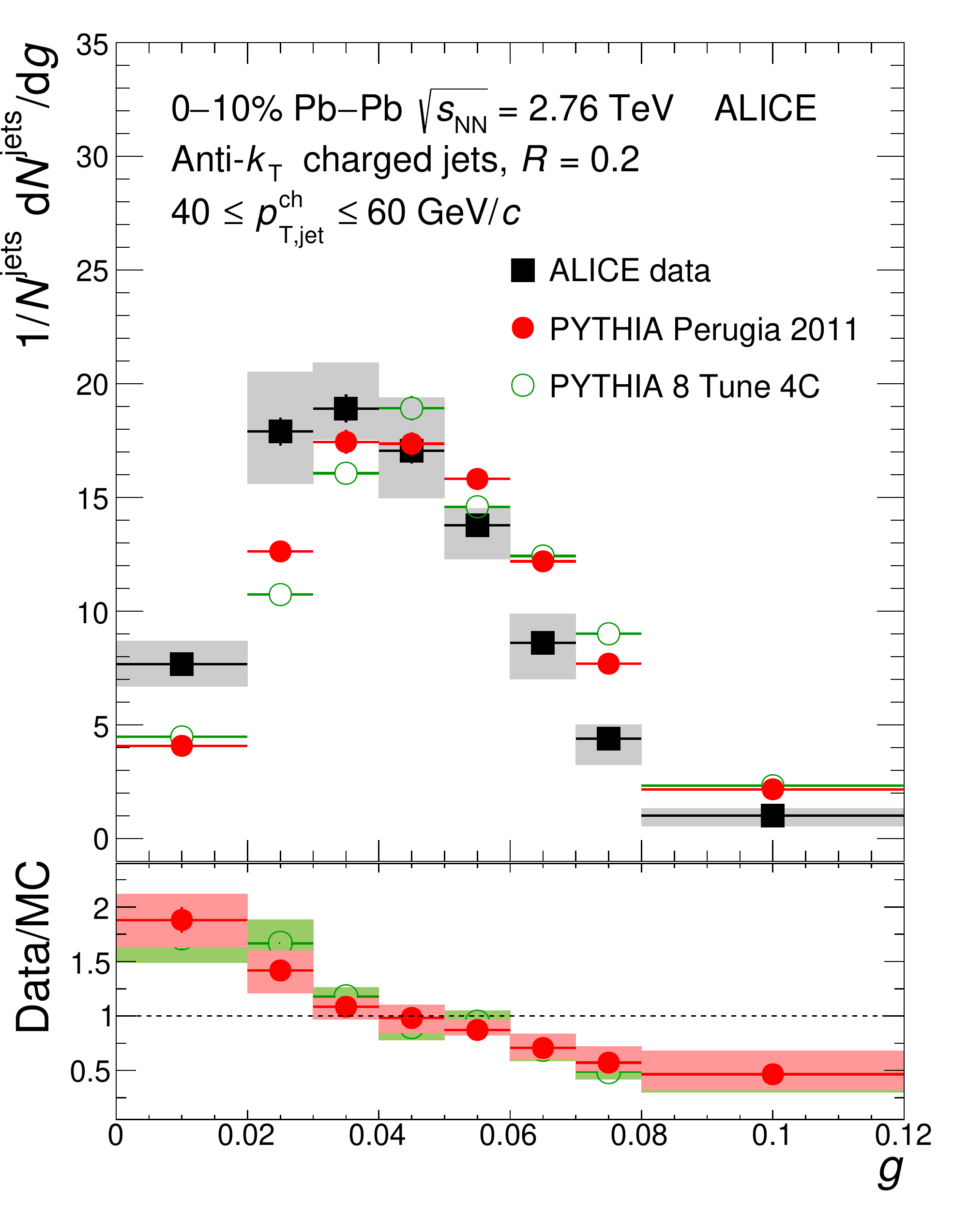}
\includegraphics[width=0.45\textwidth,height=0.45\textheight]{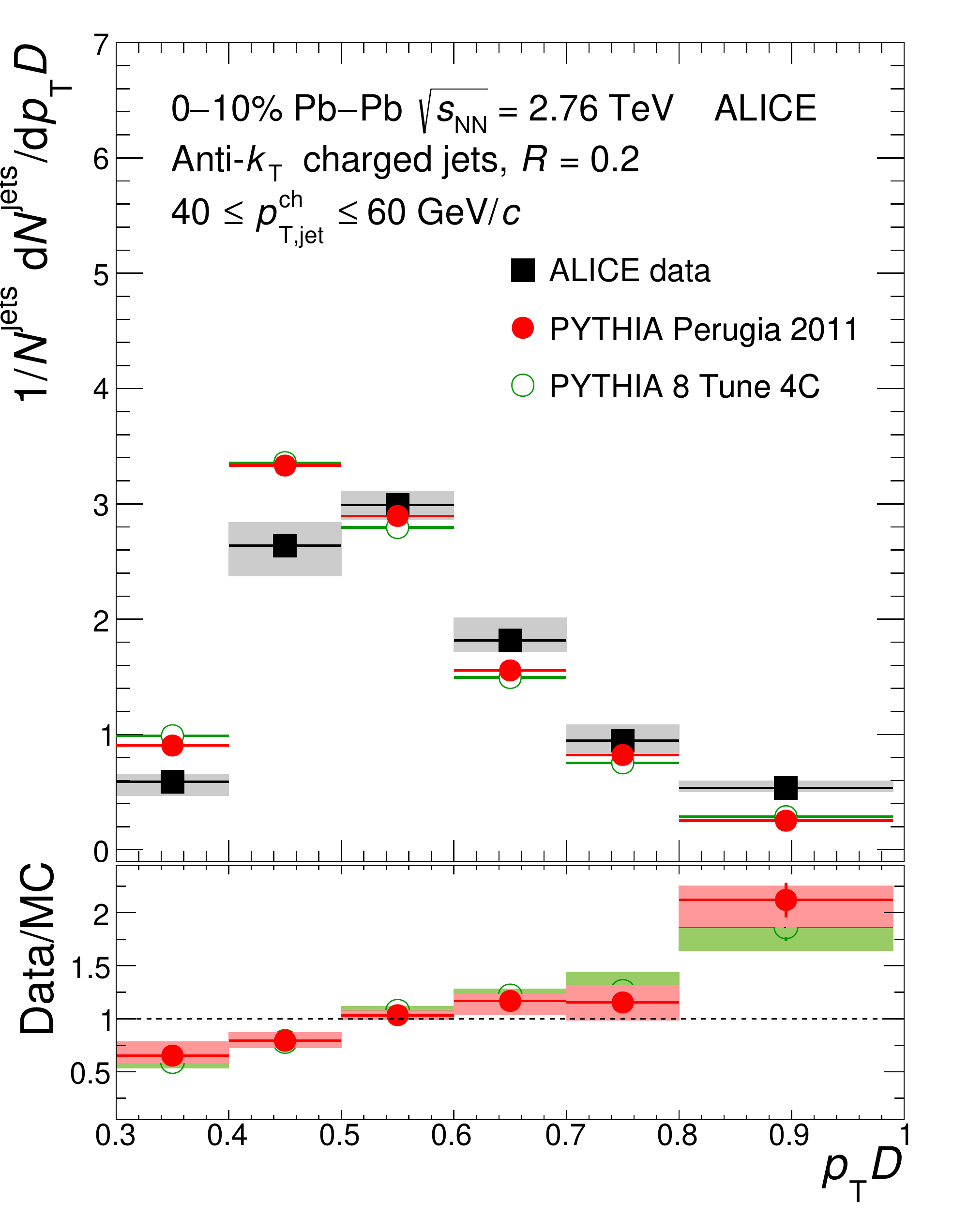}
\includegraphics[width=0.45\textwidth,height=0.45\textheight]{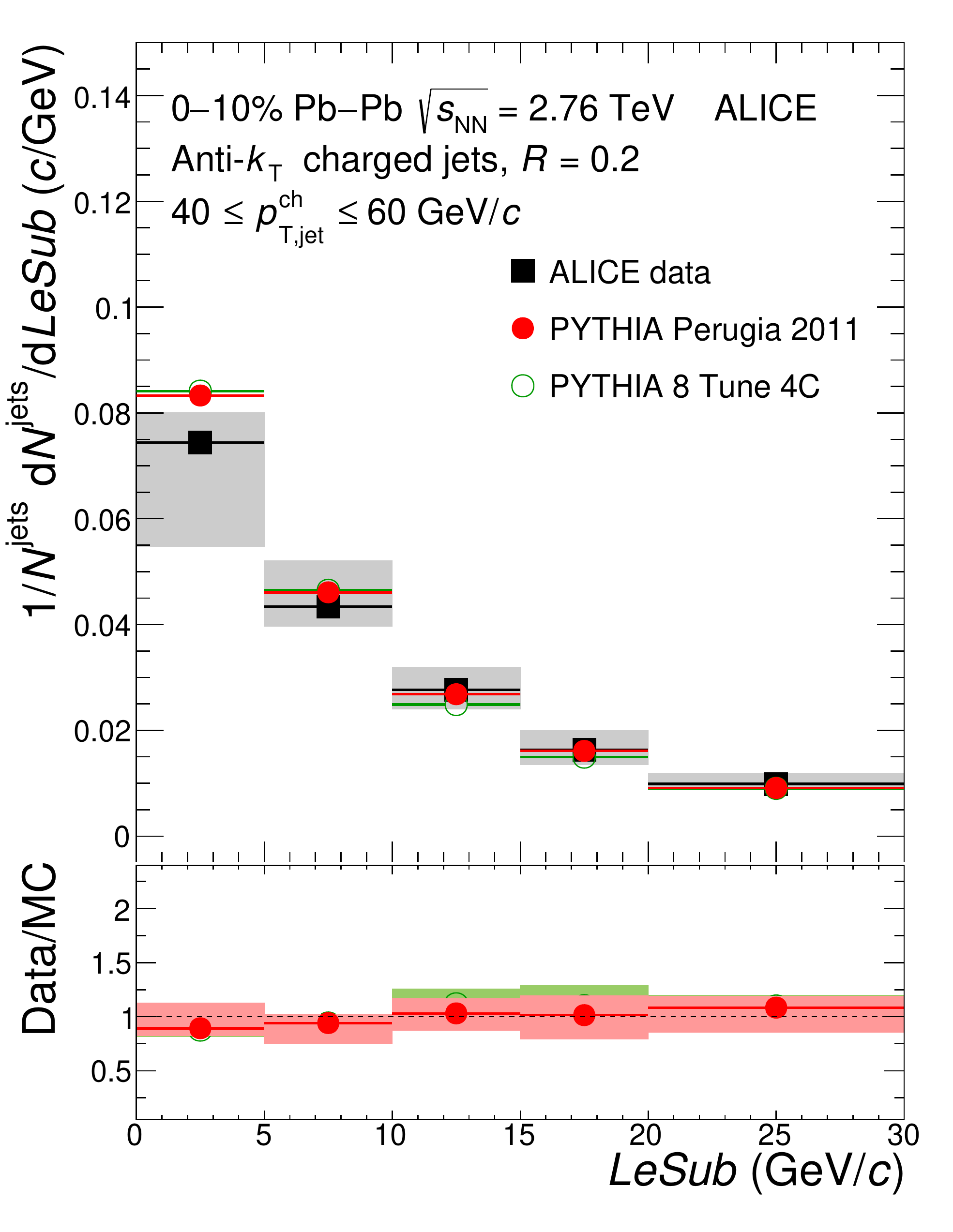}

\caption{Fully corrected jet shape distributions in $0$--$10\%$ central Pb--Pb collisions at $\sqrt{s_{\rm NN}} = 2.76$\,TeV for $R = 0.2$ in 
the range of jet $p_{\mathrm{T,jet}}^{\rm ch}$ of $40$--$60$\,GeV$/c$. The results are compared to PYTHIA. The coloured boxes represent the uncertainty on the jet shape (upper panels) and its propagation to the ratio (lower panels).}
\label{fig:ResultsPbPb}
\end{figure}

\noindent Figure~\ref{fig:ResultsPbPb} shows the fully corrected jet shape distributions in Pb--Pb
collisions at $\ensuremath{\sqrt{s_{\mathrm{NN}}}} =$ 2.76\,TeV compared to PYTHIA Perugia 2011 and PYTHIA 8 tune 4C at the same collision energy and in the
same jet $p_{\rm T}$ range of $40$--$60$\,GeV/$c$.
The radial moment (upper left plot) appears to be shifted to lower values in the measured data compared to PYTHIA. The $p_{\mathrm{T}}D$
(upper right plot) is shifted to higher values in the measured data compared to
PYTHIA. $LeSub$ (bottom) shows no indication of modifications relative
to PYTHIA. These results indicate that the fragmentation in Pb--Pb
collisions is harder and more collimated than in vacuum at the same reconstructed energy. 

The observed hardening of the fragmentation is qualitatively consistent with the observed enhancement of the high-$z$ component of the fragmentation functions of inclusive jets measured by ATLAS and CMS in Pb--Pb collisions\cite{Aaboud:2017bzv, Chatrchyan:2014ava}. More recent measurements of fragmentation functions of jets recoiling from photons at CMS \cite{Sirunyan:2018qec} do not show an enhancement at high-$z$ but rather indicate a depletion of the high-$z$ component accompanied by an enhancement of the soft modes. When the jet fragmentation is studied as a function of the photon energy in gamma-jet events, where the transverse momentum of the photon 
balances the initial parton momentum from the hard scattering to good approximation, there is no bias towards higher $Q^{2}$ in Pb--Pb relative to pp compared to the case when the recoiling jet energy is used. To quantitatively compare the different observables that select different samples of jets (inclusive vs recoil) and that are subject to different kinematic cuts, modeling within the same theoretical framework is required. 

\begin{figure}[h]

\includegraphics[width=0.33\textwidth,height=0.33\textwidth]{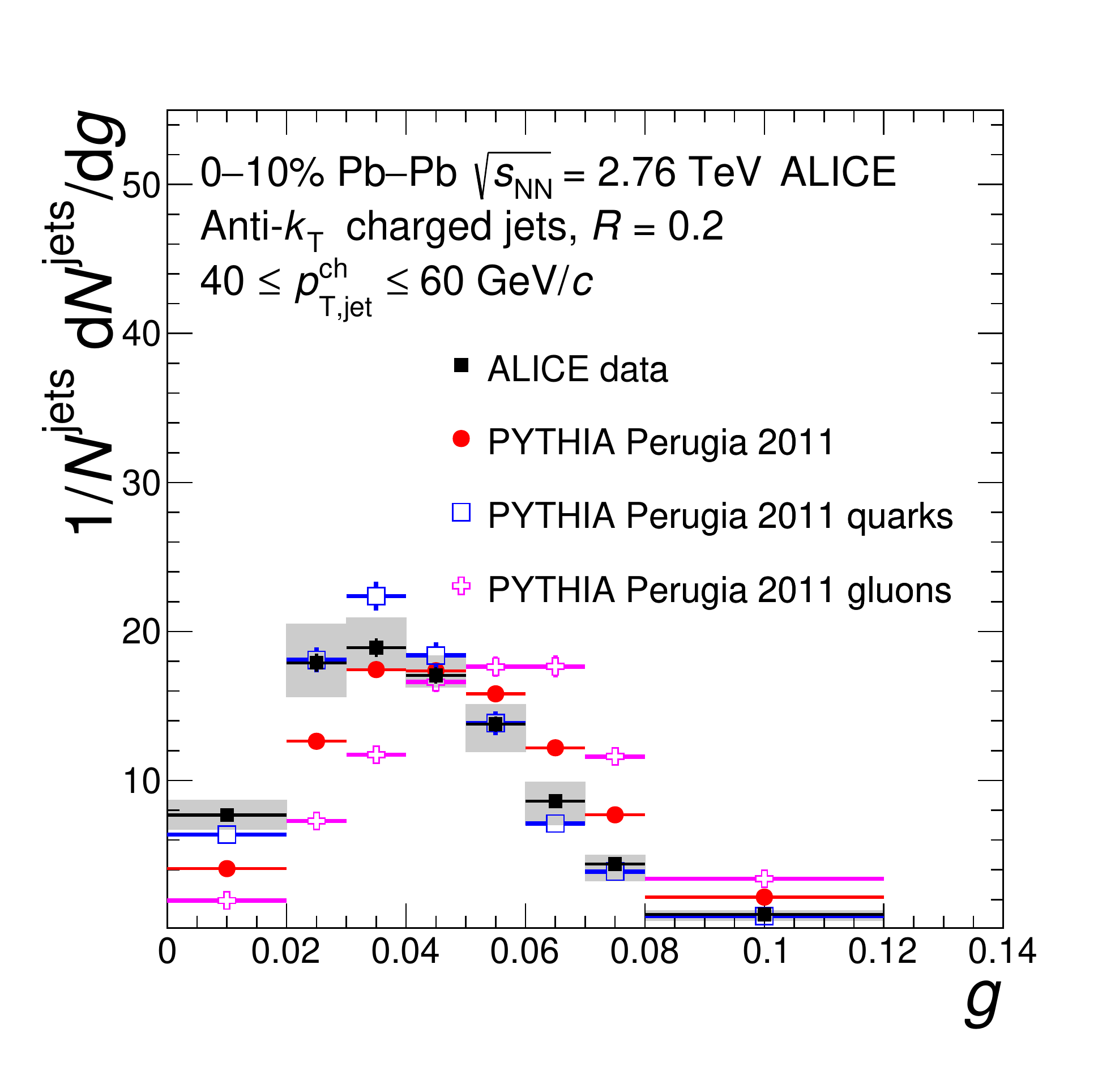}
\includegraphics[width=0.33\textwidth,height=0.33\textwidth]{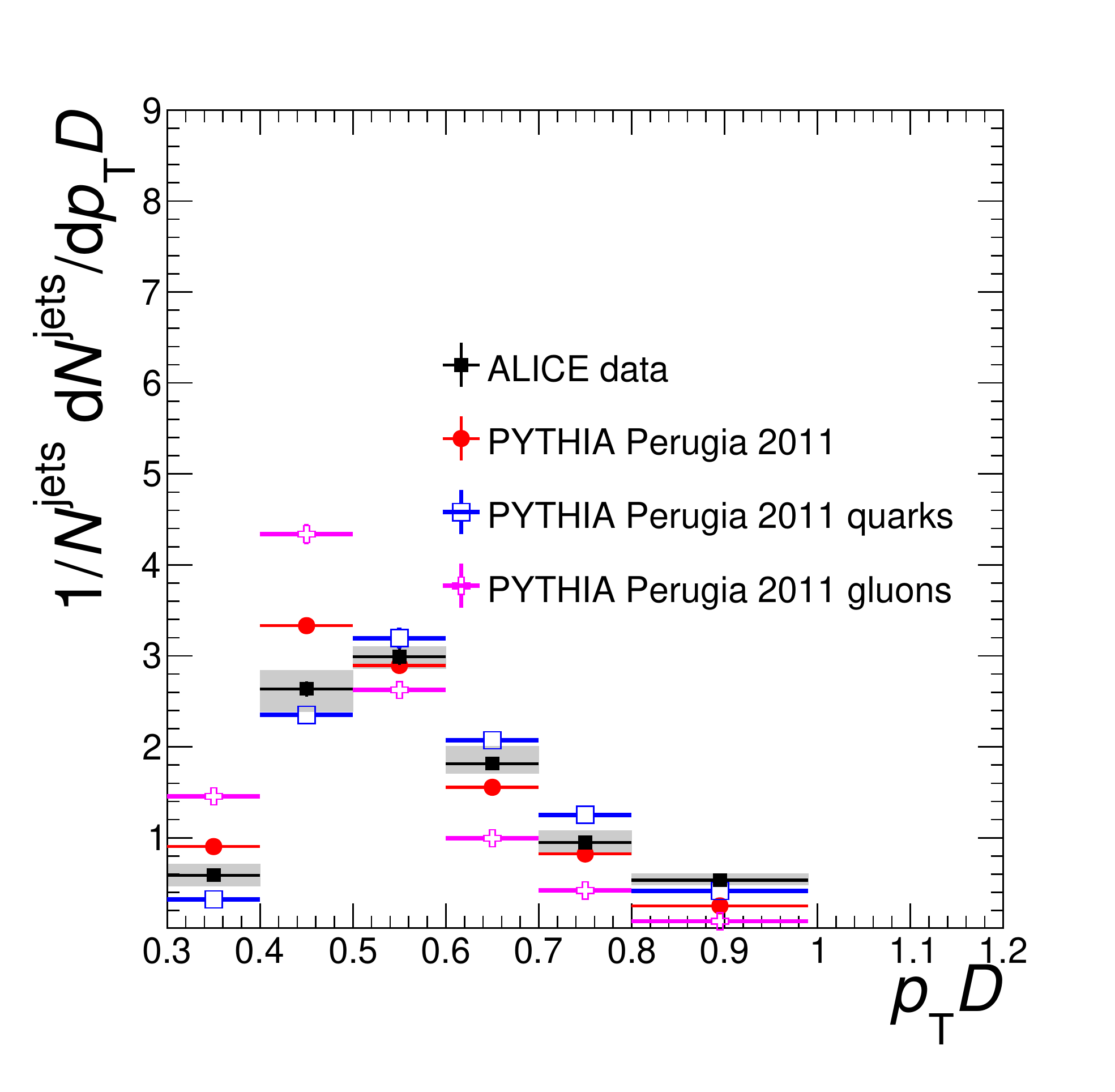}
\includegraphics[width=0.33\textwidth,height=0.33\textwidth]{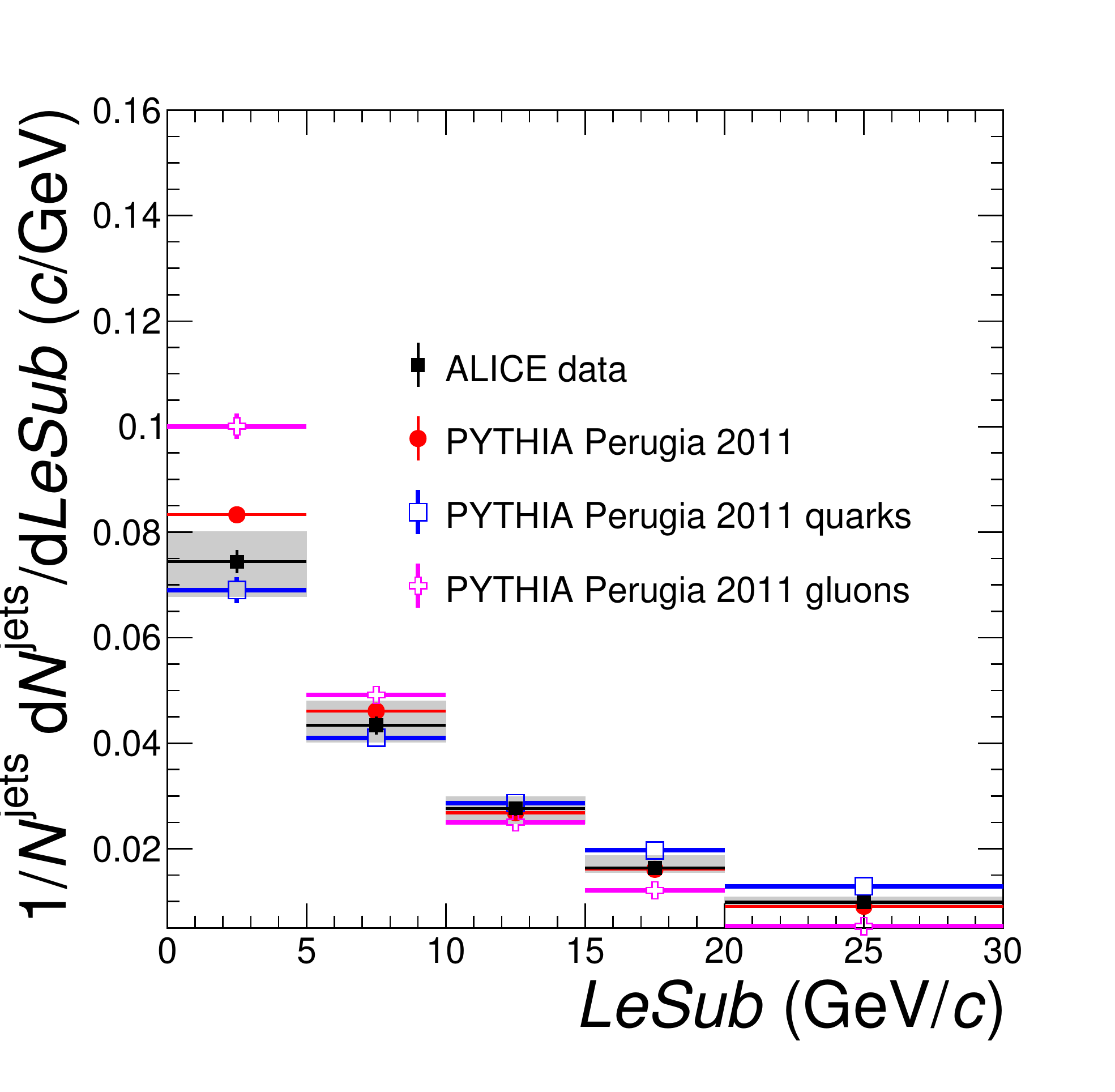}
\caption{Jet shape distributions in $0$--$10\%$ central Pb--Pb collisions at $\sqrt{s_{\rm NN}}=2.76$\,TeV for $R = 0.2$ in 
range of jet $p_{\mathrm{T,jet}}^{\rm ch}$ of $40$--$60$\,GeV$/c$ compared to quark and gluon vacuum generated jet shape distributions. The coloured boxes represent the experimental uncertainty on the jet shapes.}
\label{fig:ShapesQG}
   \end{figure}

 In Fig.~\ref{fig:ShapesQG} we compared quark and gluon vacuum jet shape distributions from PYTHIA to our data. Since quark-initiated jets radiate less, their fragmentation is harder and less broad. Gluon-initiated jets can be thought of as an approximation to modified jets in the hypothetical case where quenching accelerates the shower evolution just
by increasing the number of splittings. This scenario would lead to a
broadening and softening of the in-cone shower (see differences in the
shape between inclusive jets and gluon jets in the plot) as
opposed to the data. 
 The comparison in Fig.~\ref{fig:ShapesQG} indicates that the Pb--Pb fragmentation agrees more with a vacuum quark-like fragmentation than with a vacuum gluon-like fragmentation. It is worth noting that in the case where gluon jets interact more strongly with the medium than quark jets, their relative fractions might change for a given jet $p_{\rm T}$ in favour of more quark-initiated jets. In line with this argument, the simple toy model calculations described in Ref.~\cite{Spousta:2015fca} can explain qualitatively some aspects of the data like the hardening of the fragmentation function and $p_{\rm T}$ dependence of the jet suppression, just by using a varying quark fraction and a greater quenching for gluon jets.  

Another ingredient that might contribute to the observed differences
between jet shapes in Pb--Pb and pp collisions at the same
$p_{\rm{T,jet}}^{\rm ch}$ is that the original energy of the parton
initiating the jet shower is different in both systems. The
significant suppression of jet rates at high $p_{\rm{T}}$, $R_{\rm AA}
< 1$, suggests that the jet 
energy that is reconstructed in Pb--Pb
collisions is smaller than the 
original parton energy; this could lead to a larger virtuality of jets in Pb--Pb than in pp collisions for a given momentum. 
Let's consider the case where a fraction $X\%$ of the jet momentum is lost coherently, meaning that the jet substructure is not resolved by the medium and the jet radiates as a single colour charge~\cite{CasalderreySolana:2012ef}. Since $g$ and $p_{\rm{T}}D$ are normalized to the jet $p_{\rm{T}}$,
a simple rescaling by a momentum fraction $X\%$ of each jet constituent leaves
Eqs.~\ref{eq:angu} and~\ref{eq:ptd} unmodified. In this scenario, the
modified jet shapes, for a given reconstructed jet $p_{\rm{T}}$, are simply
the vacuum-like shapes of jets with a momentum higher by a fraction
$1/(1-X\%)$. As seen in Fig.~\ref{fig:ptdeppythia}, both $g$ and $p_{\rm{T}}D$ decrease with jet
momentum in vacuum. Our experimental results show that the $g$ distribution shifts to lower values in Pb--Pb collisions
relative to the vacuum-like one. The $p_{\rm{T}}D$ distribution, instead, increases, contrary to what is
expected from a fully coherent energy loss scenario. Following these
considerations, the medium seems to be able to resolve the jet structure at angular scales below $R=0.2$.      



\begin{figure}[h]

\includegraphics[width=0.33\textwidth,height=0.33\textwidth]{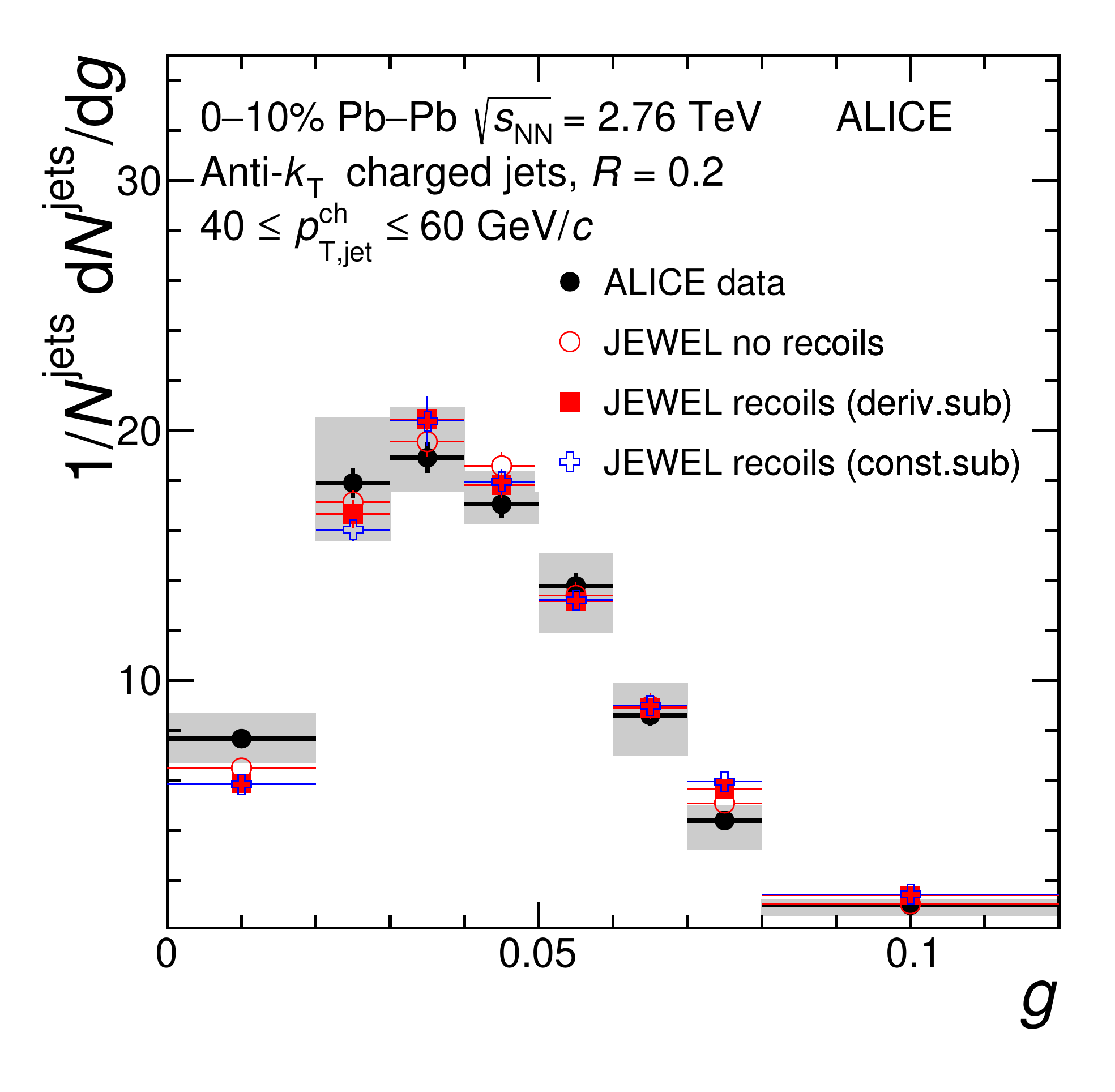}
\includegraphics[width=0.33\textwidth,height=0.33\textwidth]{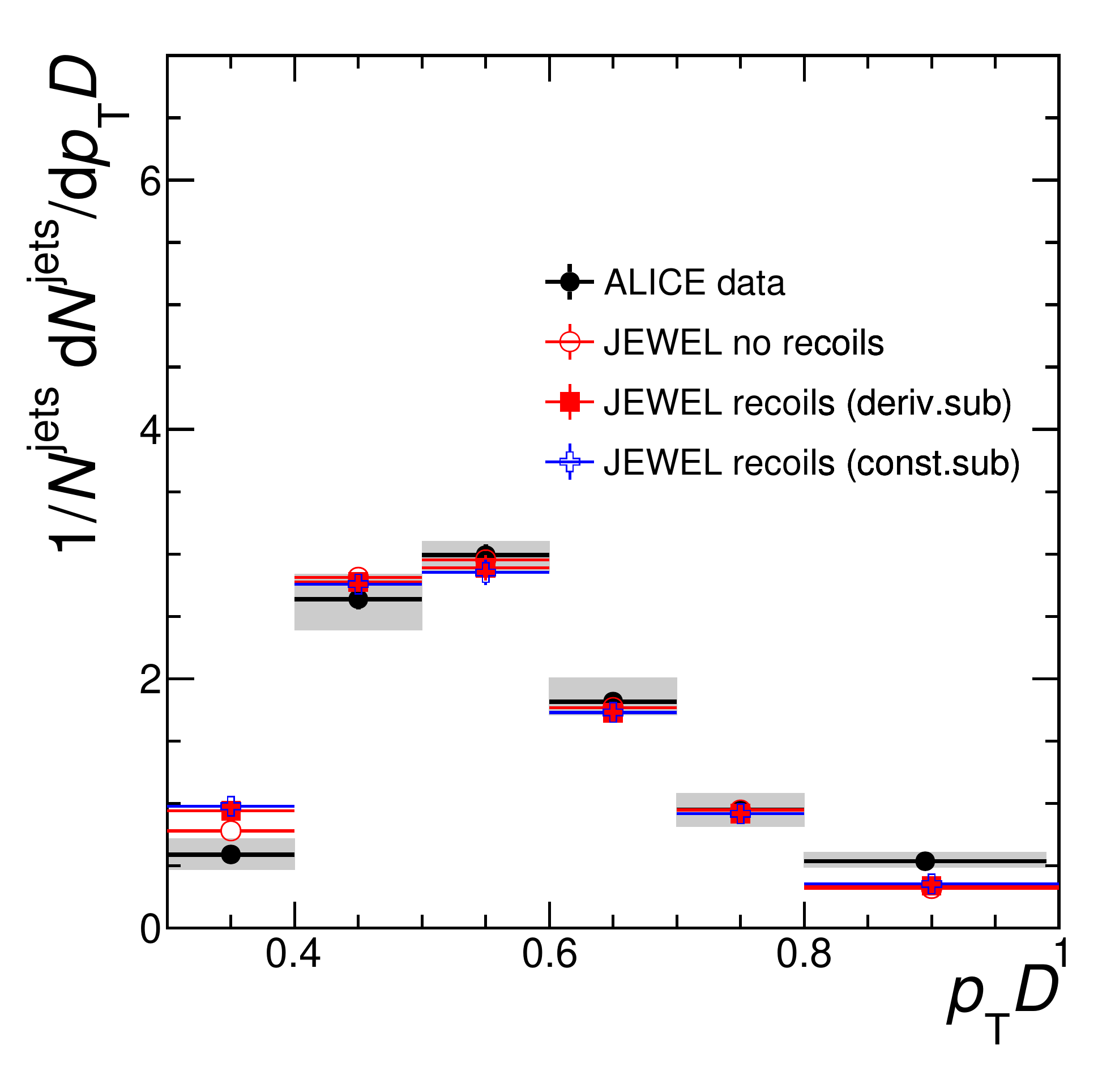}
\includegraphics[width=0.33\textwidth,height=0.33\textwidth]{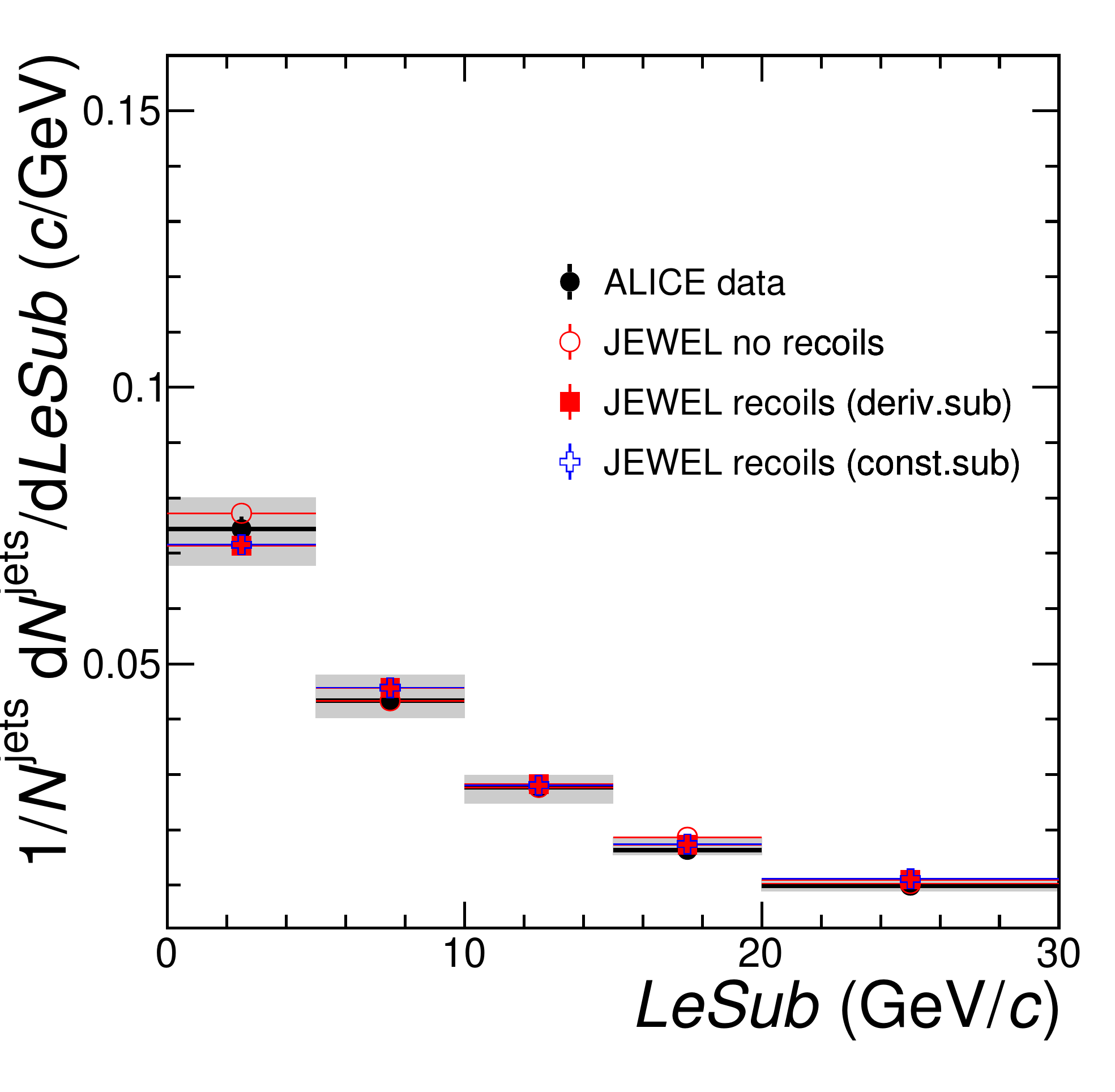}

\caption{Jet shape distributions in $0$--$10\%$ central Pb--Pb collisions at $\sqrt{s_{\rm NN}}=2.76$\,TeV for $R = 0.2$ in 
range of jet $p_{\mathrm{T,jet}}^{\rm ch}$ of $40$--$60$\,GeV$/c$ compared to JEWEL with and without recoils with different subtraction methods. The coloured boxes represent the experimental uncertainty on the jet shapes.}
\label{fig:ShapesJEWEL}
\end{figure}

We also compared our results to JEWEL calculations \cite{Zapp:2008gi}, which is a perturbative framework for jet evolution in the presence of a dense medium. The detailed description of the jet-medium interaction includes elastic scattering off medium constituents, inelastic medium-induced gluon radiation, and medium recoil. The medium recoil refers to the response of the medium to the jet. This component is a correlated background that cannot be experimentally suppressed. An extensive comparison of the model to the existing jet shapes was done (see Ref. \cite{KunnawalkamElayavalli:2017hxo}), showing that the contribution of the medium recoil to the modification of the jet shapes is large, in particular in those shapes that are most sensitive to the soft, large-angle quanta such as the jet mass or the subjet momentum imbalance $z_{\rm g}$. 
Figure \ref{fig:ShapesJEWEL} shows the measured jet shape distributions compared to JEWEL calculations. The effects of the medium recoil are small, as expected for the small considered $R$ and thus the measurement constrains the purely radiative aspects of the JEWEL shower modification. There is good agreement between the model and the data.

The ALICE measurement of the jet mass \cite{Acharya:2017goa} for jets of $R = 0.4$ showed some hints of a reduction relative to the vacuum reference. The jet mass, as discussed in Section \ref{sect:definition}, differs parametrically from the angularity only in the power of the angle dependence, so it is also sensitive to the broadening or collimation of the jet shower. Comparisons to JEWEL revealed that the effect of jet mass reduction due to energy loss is obscured by the broadening due to the medium response or recoil\cite{KunnawalkamElayavalli:2017hxo}, which contributes more strongly to jets with $R = 0.4$ than to the jet core measurements with $R = 0.2$ reported here.

 \section{Conclusions}
\label{sect:conclusions}
In this paper, the first measurement of a new set of track-based jet
shape distributions has been presented. The measurements were performed in pp collisions at
$\sqrt{s} = 7$\, TeV and in $0$--$10\%$ central Pb--Pb collisions at
$\sqrt{s_{\rm NN}} = 2.76$\,TeV in the low jet transverse momentum
interval $40 \leq p_{\mathrm{T,jet}}^{\rm{ch}} \le 60$\,GeV/$c$ and
using small jet resolution $R=$ 0.2. The full correction to particle
level and the measurement of an unbiased sample of jets with a
constituent transverse momentum cutoff of $0.15$ GeV/$c$ are key aspects of the analysis that allow exploring possible medium modifications in a wide dynamical range including soft modes. 

The jet shapes reported here probe complementary
aspects of the jet fragmentation and are used to test possible scenarios and ingredients of the theoretical description of jet quenching. 

The measurements of $g$, $p_{\mathrm{T}}D$, and $LeSub$ in pp
collisions are within 20\% in agreement with PYTHIA Perugia 2011 and PYTHIA 8 4C tunes.

In central Pb--Pb collisions, the measurements of $g$ and $p_{\mathrm{T}}D$
show that the jet core is more collimated and fragments harder than in pp collisions. The picture is qualitatively
consistent with a more quark-like jet fragmentation, suggesting e.g. either a modified fragmentation pattern of all jets or a selection on quark-like jet properties imposed by the medium for these observables. 

The role of colour coherence in the jet-medium interactions and the scales at which it dominates is a key ingredient in the characterisation of the medium. We argue that the medium-modification of $g$ and $p_{\rm{T}}D$ is not consistent with the scenario where the jets interact with the medium coherently, as single colour charges. This suggests that the medium is able to resolve the jet structure at angular scales smaller than $R=0.2$. 

Comparison to calculations using the JEWEL jet quenching model shows that the contribution of the medium response to the small-radius jets reported here is small and thus the data can constrain the effects due to energy loss, contrary to other measurements at larger $R$ where the medium recoil can obscure the radiative effects.

\newenvironment{acknowledgement}{\relax}{\relax}
\begin{acknowledgement}
\section*{Acknowledgements}

The ALICE Collaboration would like to thank all its engineers and technicians for their invaluable contributions to the construction of the experiment and the CERN accelerator teams for the outstanding performance of the LHC complex.
The ALICE Collaboration gratefully acknowledges the resources and support provided by all Grid centres and the Worldwide LHC Computing Grid (WLCG) collaboration.
The ALICE Collaboration acknowledges the following funding agencies for their support in building and running the ALICE detector:
A. I. Alikhanyan National Science Laboratory (Yerevan Physics Institute) Foundation (ANSL), State Committee of Science and World Federation of Scientists (WFS), Armenia;
Austrian Academy of Sciences and Nationalstiftung f\"{u}r Forschung, Technologie und Entwicklung, Austria;
Ministry of Communications and High Technologies, National Nuclear Research Center, Azerbaijan;
Conselho Nacional de Desenvolvimento Cient\'{\i}fico e Tecnol\'{o}gico (CNPq), Universidade Federal do Rio Grande do Sul (UFRGS), Financiadora de Estudos e Projetos (Finep) and Funda\c{c}\~{a}o de Amparo \`{a} Pesquisa do Estado de S\~{a}o Paulo (FAPESP), Brazil;
Ministry of Science \& Technology of China (MSTC), National Natural Science Foundation of China (NSFC) and Ministry of Education of China (MOEC) , China;
Ministry of Science and Education, Croatia;
Ministry of Education, Youth and Sports of the Czech Republic, Czech Republic;
The Danish Council for Independent Research | Natural Sciences, the Carlsberg Foundation and Danish National Research Foundation (DNRF), Denmark;
Helsinki Institute of Physics (HIP), Finland;
Commissariat \`{a} l'Energie Atomique (CEA) and Institut National de Physique Nucl\'{e}aire et de Physique des Particules (IN2P3) and Centre National de la Recherche Scientifique (CNRS), France;
Bundesministerium f\"{u}r Bildung, Wissenschaft, Forschung und Technologie (BMBF) and GSI Helmholtzzentrum f\"{u}r Schwerionenforschung GmbH, Germany;
General Secretariat for Research and Technology, Ministry of Education, Research and Religions, Greece;
National Research, Development and Innovation Office, Hungary;
Department of Atomic Energy Government of India (DAE), Department of Science and Technology, Government of India (DST), University Grants Commission, Government of India (UGC) and Council of Scientific and Industrial Research (CSIR), India;
Indonesian Institute of Science, Indonesia;
Centro Fermi - Museo Storico della Fisica e Centro Studi e Ricerche Enrico Fermi and Istituto Nazionale di Fisica Nucleare (INFN), Italy;
Institute for Innovative Science and Technology , Nagasaki Institute of Applied Science (IIST), Japan Society for the Promotion of Science (JSPS) KAKENHI and Japanese Ministry of Education, Culture, Sports, Science and Technology (MEXT), Japan;
Consejo Nacional de Ciencia (CONACYT) y Tecnolog\'{i}a, through Fondo de Cooperaci\'{o}n Internacional en Ciencia y Tecnolog\'{i}a (FONCICYT) and Direcci\'{o}n General de Asuntos del Personal Academico (DGAPA), Mexico;
Nederlandse Organisatie voor Wetenschappelijk Onderzoek (NWO), Netherlands;
The Research Council of Norway, Norway;
Commission on Science and Technology for Sustainable Development in the South (COMSATS), Pakistan;
Pontificia Universidad Cat\'{o}lica del Per\'{u}, Peru;
Ministry of Science and Higher Education and National Science Centre, Poland;
Korea Institute of Science and Technology Information and National Research Foundation of Korea (NRF), Republic of Korea;
Ministry of Education and Scientific Research, Institute of Atomic Physics and Romanian National Agency for Science, Technology and Innovation, Romania;
Joint Institute for Nuclear Research (JINR), Ministry of Education and Science of the Russian Federation and National Research Centre Kurchatov Institute, Russia;
Ministry of Education, Science, Research and Sport of the Slovak Republic, Slovakia;
National Research Foundation of South Africa, South Africa;
Centro de Aplicaciones Tecnol\'{o}gicas y Desarrollo Nuclear (CEADEN), Cubaenerg\'{\i}a, Cuba and Centro de Investigaciones Energ\'{e}ticas, Medioambientales y Tecnol\'{o}gicas (CIEMAT), Spain;
Swedish Research Council (VR) and Knut \& Alice Wallenberg Foundation (KAW), Sweden;
European Organization for Nuclear Research, Switzerland;
National Science and Technology Development Agency (NSDTA), Suranaree University of Technology (SUT) and Office of the Higher Education Commission under NRU project of Thailand, Thailand;
Turkish Atomic Energy Agency (TAEK), Turkey;
National Academy of  Sciences of Ukraine, Ukraine;
Science and Technology Facilities Council (STFC), United Kingdom;
National Science Foundation of the United States of America (NSF) and United States Department of Energy, Office of Nuclear Physics (DOE NP), United States of America.
\end{acknowledgement}

\bibliographystyle{utphys}
\bibliography{references.bib}



\newpage
\appendix
\section{The ALICE Collaboration}
\label{app:collab}

\begingroup
\small
\begin{flushleft}
S.~Acharya$^{\rm 139}$, 
F.T.-.~Acosta$^{\rm 20}$, 
D.~Adamov\'{a}$^{\rm 93}$, 
A.~Adler$^{\rm 74}$, 
J.~Adolfsson$^{\rm 80}$, 
M.M.~Aggarwal$^{\rm 98}$, 
G.~Aglieri Rinella$^{\rm 34}$, 
M.~Agnello$^{\rm 31}$, 
N.~Agrawal$^{\rm 48}$, 
Z.~Ahammed$^{\rm 139}$, 
S.U.~Ahn$^{\rm 76}$
% ...

\bigskip

\bigskip 

\textbf{\Large Affiliation Notes}

\bigskip 

% ...
$^{\rm V}$ Also at: Institute of Theoretical Physics, University of Wroclaw, Poland\\

\bigskip

\bigskip 

\textbf{\Large Collaboration Institutes}

\bigskip 

% ...
$^{20}$ Department of Physics, University of California, Berkeley, California, United States\\
% ...
$^{31}$ Dipartimento DISAT del Politecnico and Sezione INFN, Turin, Italy\\
% ...
$^{34}$ European Organization for Nuclear Research (CERN), Geneva, Switzerland\\
% ...
$^{48}$ Indian Institute of Technology Bombay (IIT), Mumbai, India\\
% ...
$^{74}$ Johann-Wolfgang-Goethe Universit\"{a}t Frankfurt Institut f\"{u}r Informatik, Fachbereich Informatik und Mathematik, Frankfurt, Germany\\
% ...
$^{76}$ Korea Institute of Science and Technology Information, Daejeon, Republic of Korea\\
% ...
$^{80}$ Lund University Department of Physics, Division of Particle Physics, Lund, Sweden\\
% ...
$^{93}$ Nuclear Physics Institute of the Czech Academy of Sciences, \v{R}e\v{z} u Prahy, Czech Republic\\
% ...
$^{98}$ Physics Department, Panjab University, Chandigarh, India\\
% ...
$^{139}$ Variable Energy Cyclotron Centre, Homi Bhabha National Institute, Kolkata, India\\
% ...

\bigskip 

\end{flushleft}

\end{document}